\documentclass[preprint2]{emulateapj} 
\usepackage{amsmath}
\usepackage{natbib}
\usepackage{graphicx}
\usepackage{epsf}
\usepackage{color}
\input{epsf}
\usepackage{epsfig}
\usepackage{rotating}
\usepackage{afterpage}
\DeclareGraphicsExtensions{.jpg,.pdf,.png,.eps,.ps}
\graphicspath{{FIGURES/}}
\usepackage{lipsum}

\newcommand{\Tcmb}{\mbox{$T_{\mbox{\tiny CMB}}$}}

\newcommand{\hmsun}{h^{-1} \; M_{\odot}}
\newcommand{\LCDM}{\mbox{$\Lambda$}CDM}
\newcommand{\lcdm}{\mbox{$\Lambda$}CDM}
\newcommand{\planck}{{\it Planck}}
\newcommand{\wmap}{{\it WMAP}}
\newcommand{\herschel}{{\it Herschel}}
\newcommand{\spitzer}{{\it Spitzer}}

\newcommand{\sqdeg}{\mbox{deg$^2$}}
\newcommand{\chisq}{\ensuremath{\chi^2}}
\newcommand{\delchisq}{\ensuremath{\Delta\chi^2}}

\newcommand{\ltsima}{$\; \buildrel < \over \sim \;$}
\newcommand{\ltsim}{\lower.5ex\hbox{\ltsima}}

\newcommand{\sptsz}{SPT-SZ}
\newcommand{\summnu}{\ensuremath{\sum m_\nu}}
\newcommand{\ho}{\ensuremath{\rm H_{0}}}
\newcommand{\neff}{\ensuremath{N_{\rm eff}}}

\newcommand{\uksq}{\ensuremath{\mu {\rm K}^2}}
\newcommand{\dksz}{\ensuremath{D_{3000}^{\rm kSZ}}}
\newcommand{\dtsz}{\ensuremath{D_{3000}^{\rm tSZ}}}
\newcommand{\zmid}{\ensuremath{z_\mathrm{mid}}}
\newcommand{\zend}{\ensuremath{z_\mathrm{end}}}
\newcommand\comment[1]{}

\def\xhii{{\bar x}_e}

\newcommand{\degsq}{deg$^2$}

\def\Berkeley{1}
\def\MPE{2}
\def\Melbourne{3}
\def\UChicago{4}
\def\AAUChicago{5}
\def\KICPChicago{6}
\def\FNAL{7}
\def\PhysicsUChicago{8}
\def\Argonne{9}
\def\EFIChicago{10}
\def\NIST{11}
\def\Caltech{12}
\def\McGill{13}
\def\Colorado{14}
\def\Davis{15}
\def\StanfordKPAC{16}
\def\StanfordPhys{17}
\def\LBNL{18}
\def\Michigan{19}
\def\Munich{20}
\def\ExcellenceCluster{21}
\def\CaseWestern{22}
\def\Minnesota{23}
\def\ArtInstChicago{24}
\def\CfA{25}
\def\Dunlap{26}
\def\Toronto{27}
\def\illast{28}
\def\illphy{29}
\def\BCCP{30}
\begin{document}

\title{A measurement of secondary cosmic microwave background anisotropies from the 2500-square-degree \sptsz{} survey}

\author{
  E.~M.~George\altaffilmark{\Berkeley,\MPE},
  C.~L.~Reichardt\altaffilmark{\Berkeley,\Melbourne},
  K.~A.~Aird\altaffilmark{\UChicago},
  B.~A.~Benson\altaffilmark{\AAUChicago,\KICPChicago,\FNAL},
  L.~E.~Bleem\altaffilmark{\KICPChicago,\PhysicsUChicago,\Argonne},
  J.~E.~Carlstrom\altaffilmark{\AAUChicago,\KICPChicago,\PhysicsUChicago,\Argonne,\EFIChicago},
  C.~L.~Chang\altaffilmark{\KICPChicago,\Argonne,\EFIChicago},
  H-M.~Cho\altaffilmark{\NIST},
  T.~M.~Crawford\altaffilmark{\AAUChicago,\KICPChicago}
  A.~T.~Crites\altaffilmark{\AAUChicago,\KICPChicago,\Caltech},
  T.~de~Haan\altaffilmark{\McGill},
  M.~A.~Dobbs\altaffilmark{\McGill},
  J.~Dudley\altaffilmark{\McGill},
  N.~W.~Halverson\altaffilmark{\Colorado},
  N.~L.~Harrington\altaffilmark{\Berkeley},
  G.~P.~Holder\altaffilmark{\McGill},
  W.~L.~Holzapfel\altaffilmark{\Berkeley},
  Z.~Hou\altaffilmark{\Davis},
  J.~D.~Hrubes\altaffilmark{\UChicago},
  R.~Keisler\altaffilmark{\KICPChicago,\PhysicsUChicago,\StanfordKPAC,\StanfordPhys},
  L.~Knox\altaffilmark{\Davis},
  A.~T.~Lee\altaffilmark{\Berkeley,\LBNL},
  E.~M.~Leitch\altaffilmark{\AAUChicago,\KICPChicago},
  M.~Lueker\altaffilmark{\Caltech},
  D.~Luong-Van\altaffilmark{\UChicago},
  J.~J.~McMahon\altaffilmark{\Michigan},
  J.~Mehl\altaffilmark{\KICPChicago,\Argonne},
  S.~S.~Meyer\altaffilmark{\AAUChicago,\KICPChicago,\PhysicsUChicago,\EFIChicago},
  M.~Millea\altaffilmark{\Davis},
  L.~M.~Mocanu\altaffilmark{\AAUChicago,\KICPChicago},
  J.~J.~Mohr\altaffilmark{\MPE,\Munich,\ExcellenceCluster},
  T.~E.~Montroy\altaffilmark{\CaseWestern},
  S.~Padin\altaffilmark{\AAUChicago,\KICPChicago,\Caltech},
  T.~Plagge\altaffilmark{\AAUChicago,\KICPChicago},
  C.~Pryke\altaffilmark{\Minnesota},
  J.~E.~Ruhl\altaffilmark{\CaseWestern},
  K.~K.~Schaffer\altaffilmark{\KICPChicago,\EFIChicago,\ArtInstChicago},
  L.~Shaw\altaffilmark{\McGill},
  E.~Shirokoff\altaffilmark{\AAUChicago,\KICPChicago}, 
  H.~G.~Spieler\altaffilmark{\LBNL},
  Z.~Staniszewski\altaffilmark{\CaseWestern},
  A.~A.~Stark\altaffilmark{\CfA},
  K.~T.~Story\altaffilmark{\KICPChicago,\PhysicsUChicago},
  A.~van~Engelen\altaffilmark{\McGill},
  K.~Vanderlinde\altaffilmark{\Dunlap,\Toronto},
  J.~D.~Vieira\altaffilmark{\illast, \illphy},
  R.~Williamson\altaffilmark{\AAUChicago,\KICPChicago}, and
  O.~Zahn\altaffilmark{\BCCP}
}

\altaffiltext{\Berkeley}{Department of Physics, University of California, Berkeley, CA, USA 94720}
\altaffiltext{\MPE}{Max-Planck-Institut f\"{u}r extraterrestrische Physik, 85748 Garching, Germany}
\altaffiltext{\Melbourne}{School of Physics, University of Melbourne, Parkville, VIC 3010, Australia}
\altaffiltext{\UChicago}{University of Chicago, Chicago, IL, USA 60637}
\altaffiltext{\AAUChicago}{Department of Astronomy and Astrophysics, University of Chicago, Chicago, IL, USA 60637}
\altaffiltext{\KICPChicago}{Kavli Institute for Cosmological Physics, University of Chicago, Chicago, IL, USA 60637}
\altaffiltext{\FNAL}{Fermi National Accelerator Laboratory, Batavia, IL 60510-0500, USA}
\altaffiltext{\PhysicsUChicago}{Department of Physics, University of Chicago, Chicago, IL, USA 60637}
\altaffiltext{\Argonne}{Argonne National Laboratory, Argonne, IL, USA 60439}
\altaffiltext{\EFIChicago}{Enrico Fermi Institute, University of Chicago, Chicago, IL, USA 60637}
\altaffiltext{\NIST}{NIST Quantum Devices Group, Boulder, CO, USA 80305}
\altaffiltext{\Caltech}{California Institute of Technology, Pasadena, CA, USA 91125}
\altaffiltext{\McGill}{Department of Physics, McGill University, Montreal, Quebec H3A 2T8, Canada}
\altaffiltext{\Colorado}{Department of Astrophysical and Planetary Sciences and Department of Physics, University of Colorado, Boulder, CO, USA 80309}
\altaffiltext{\Davis}{Department of Physics, University of California, Davis, CA, USA 95616}
\altaffiltext{\StanfordKPAC}{Kavli Institute for Particle Astrophysics and Cosmology, Stanford University, 452 Lomita Mall, Stanford, CA 94305}
\altaffiltext{\StanfordPhys}{Department of Physics, Stanford University, 382 Via Pueblo Mall, Stanford, CA 94305}
\altaffiltext{\LBNL}{Physics Division, Lawrence Berkeley National Laboratory, Berkeley, CA, USA 94720}
\altaffiltext{\Michigan}{Department of Physics, University of Michigan, Ann  Arbor, MI, USA 48109}
\altaffiltext{\Munich}{Department of Physics, Ludwig-Maximilians-Universit\"{a}t, 81679 M\"{u}nchen, Germany}
\altaffiltext{\ExcellenceCluster}{Excellence Cluster Universe, 85748 Garching, Germany}
\altaffiltext{\CaseWestern}{Physics Department, Center for Education and Research in Cosmology and Astrophysics, Case Western Reserve University,Cleveland, OH, USA 44106}
\altaffiltext{\Minnesota}{Department of Physics, University of Minnesota, Minneapolis, MN, USA 55455}
\altaffiltext{\ArtInstChicago}{Liberal Arts Department, School of the Art Institute of Chicago, Chicago, IL, USA 60603}
\altaffiltext{\CfA}{Harvard-Smithsonian Center for Astrophysics, Cambridge, MA, USA 02138}
\altaffiltext{\Dunlap}{Dunlap Institute for Astronomy \& Astrophysics, University of Toronto, 50 St George St, Toronto, ON, M5S 3H4, Canada}
\altaffiltext{\Toronto}{Department of Astronomy \& Astrophysics, University of Toronto, 50 St George St, Toronto, ON, M5S 3H4, Canada}
\altaffiltext{\illast}{Astronomy Department, University of Illinois at Urbana-Champaign, 1002 W.\ Green Street, Urbana, IL 61801, USA}
\altaffiltext{\illphy}{Department of Physics, University of Illinois Urbana-Champaign, 1110 W.\ Green Street, Urbana, IL 61801, USA}
\altaffiltext{\BCCP}{Berkeley Center for Cosmological Physics, Department of Physics, University of California, and Lawrence Berkeley National Laboratory, Berkeley, CA, USA 94720}

\email{lizinvt@berkeley.edu}
 
\begin{abstract}
We present measurements of secondary cosmic microwave background (CMB) anisotropies and cosmic infrared background (CIB) fluctuations using data from the South Pole Telescope (SPT) covering the complete 2540 deg$^2$ \sptsz{} survey area. 
Data in the three \sptsz{} frequency bands centered at 95, 150, and 220\,GHz, are used to produce six angular power spectra (three single-frequency auto-spectra and three cross-spectra) covering the multipole range $2000 < \ell < 11000$ (angular scales $5^\prime \gtrsim \theta \gtrsim 1^\prime$).
These are the most precise measurements of the angular power spectra at $\ell > 2500$  at these frequencies. 
The main contributors to the power spectra at these angular scales and frequencies are the primary CMB, CIB, thermal and kinematic Sunyaev-Zel'dovich effects (tSZ and kSZ), and radio galaxies.
We include a constraint on the tSZ power from a measurement of the tSZ bispectrum from 800 deg$^2$ of the \sptsz{} survey.
We measure the tSZ power at 143\, GHz to be $D^{\rm tSZ}_{3000} = 4.08^{+0.58}_{-0.67}\,\uksq{}$ and the kSZ power to be $D^{\rm kSZ}_{3000} = 2.9 \pm 1.3\, \uksq{}$. 
The data prefer positive kSZ power at 98.1\% CL. 
We measure a correlation coefficient of $\xi = 0.113^{+0.057}_{-0.054}$ between sources of tSZ and CIB power, with $\xi < 0$ disfavored at a confidence level of 99.0\%. 
The constraint on kSZ power can be interpreted as an upper limit on the duration of reionization.
When the post-reionization homogeneous kSZ signal is accounted for, we find an upper limit on the duration $\Delta z < $ 5.4\, at 95\% CL.

\end{abstract}


\bigskip\bigskip

\section{Introduction}
\setcounter{footnote}{0}
The sensitivity and angular resolution of current cosmic microwave background (CMB) experiments make it possible to study interactions between primary CMB photons and cosmic structure---so-called secondary CMB anisotropies---at small angular scales, and the cosmic infrared background (CIB) at the long-wavelength end of its spectrum.
With unprecedented precision and multi-frequency data, the current generation of high resolution instruments allow for the separation of individual contributions to the power spectra, and can provide tight constraints on the associated cosmological and astrophysical model parameters.  
These include the amplitude of local structure, the duration of the reionization epoch, and the correlation 
between dust and ionized gas in galaxy clusters.

\subsection{Expected signals}
\label{sec:intro_sig}
The most prominent secondary CMB anisotropies at few-arcminute scales (multipole number $\ell \sim 3000$)
are expected to be the 
thermal and kinematic Sunyaev-Zel'dovich (SZ) effects \citep{sunyaev72,sunyaev80b}. 
The SZ effects arise from the scattering of primary CMB photons from free electrons, both those distributed throughout the reionized (and reionizing) intergalactic medium and those concentrated in the hot gas in clusters of galaxies. 
The SZ power spectra depend sensitively on both the growth of
structure and the details of cosmic reionization.

When primary CMB photons inverse Compton scatter from electrons with a bulk flow velocity $\vec v$, the photons are Doppler shifted to higher or lower frequencies depending on the orientation of $\vec v$ with respect to the line of sight.
This kinematic SZ (kSZ) effect has a black-body spectrum, and the contribution from each volume element scales as $(v/c) n_e$ where $c$ is the speed of light and $n_e$ is the density of free electrons.
In massive galaxy clusters, photons scattering from hot electrons tend to gain energy, distorting the primary CMB black-body spectrum.
 This is known as the thermal SZ (tSZ) effect and leads to fewer observed photons below 217\,GHz and more above 217\,GHz, with a contribution from each volume element scaling as  $(k_B T_e/m_ec^2)n_e$, where $m_e$ is the mass of the electron and $T_e$ is the temperature of the electrons.  
The different frequency scalings of the tSZ and kSZ effects allows for their simultaneous determination in multi-frequency observations.

The tSZ anisotropy power depends strongly on the normalization of the matter power spectrum, commonly
parametrized by the RMS of the $z=0$ linear mass distribution on 8 Mpc/$h$ scales,
$\sigma_8$ \citep[e.g.,][]{komatsu02}. This steep scaling 
(see Section \ref{sec:tsz} for details) means that a reasonably precise measurement of the tSZ amplitude has the
statistical power to tightly constrain  $\sigma_8$.
A significant challenge in modeling the tSZ power spectrum is that it draws power from a  wide range of cluster masses and redshifts. 
Uncertainties about non-gravitational heating effects complicate models of low-mass clusters, while 
models of high-redshift clusters suffer from limited observational data.
Different prescriptions for cluster gas physics in recent models lead to differences in power of 
up to 50\% for the same cosmology \citep{sehgal10,shaw10,battaglia12,mccarthy14}.

The post-reionization kSZ power spectrum is expected to be simpler to model than the tSZ spectrum because it is not weighted by the gas temperature, and thus depends less on non-linear physics in dense halos. 
The details of cosmic reionization, however, are crucial to modeling the total kSZ power spectrum, and this epoch is not yet well understood.  
In the standard picture of cosmic reionization, ionized bubbles form around the first stars, galaxies, and quasars. 
These bubbles eventually overlap and merge, leading to a fully ionized universe. 
Proper motion of these ionized bubbles generates angular anisotropy through the kSZ effect.
The amplitude of this ``patchy'' kSZ power depends primarily on the duration of reionization, and its shape primarily on the distribution of bubble sizes; both features also depend less strongly on the average redshift of reionization \citep{gruzinov98,knox98}.
Constraints on the kSZ power spectrum can therefore lead to interesting constraints on the epoch of reionization \citep{mortonson10, zahn12,mesinger12}.

The sky power at millimeter wavelengths and scales smaller than a few arcminutes ($\ell \gtrsim 3000$) is dominated by the signal from bright extragalactic sources \citep{vieira10,marriage11a,planck11-13}, mostly synchrotron-dominated active galactic nuclei (AGN), with additional contributions from nearby infrared galaxies and rare, strongly lensed, dusty, star-forming galaxies (DSFGs, \citealt{negrello10,vieira13}).
With these identifiable sources removed, the remaining power is dominated by fainter DSFGs that contribute the bulk of the CIB \citep{lagache05,casey14}. 
These DSFGs are important both as a foreground for CMB experiments and as a tool to understand the history of star and galaxy formation \citep[e.g.,][]{bond86,bond91b,knox01,madau14}.  
The power from DSFGs can be separated into a component with no angular clustering (the ``Poisson'' component)
and an angularly clustered component.
By definition these two components have distinct angular scale dependencies.  
Multi-frequency measurements of CIB power can determine the spectral index 
of both the clustered and Poisson components of the CIB anisotropy, providing constraints on the redshift distributions
and dust temperatures for each component. \citep{addison12a}

Both the tSZ and CIB are tracers of the underlying dark matter distribution, and thus we expect the two 
signals to be correlated \citep{addison12b}. This correlation is itself an interesting probe of cluster physics and evolution, but it complicates the separation of individual components in multi-frequency measurements \citep{reichardt12b, dunkley13}.

\subsection{Previous measurements}
\label{sec:intro_prevmeas}
The first robust measurement of the amplitude of the SZ power spectrum  was presented by \citet[hereafter L10]{lueker10} using 100 \degsq{} of data from the South Pole Telescope (SPT) in the angular multipole range $(2000 < \ell < 9500)$. 
Improved SPT measurements of the power spectrum have since been reported by \citet{shirokoff11} and \citet{reichardt12b} (hereafter S11 and R12) using 200 \degsq\, and 800 \degsq\, of data respectively. 
R12 measured the tSZ and kSZ power at $\ell = 3000$ to $D^{\rm tSZ}_{3000} = 3.26\pm1.06\,\mu{\rm K}^2$ and $D^{\rm kSZ}_{3000} < 6.7\,\mu{\rm K}^2$ (95\% CL) at 153\,GHz. 
The Atacama Cosmology Telescope (ACT) collaboration has also measured the high-$\ell$ power spectrum \citep{das11b, das14} with multi-frequency band powers covering the multipole range from $(1540<\ell<9440)$. 
\citet{dunkley13} use the power spectra presented in \citet{das14} to measure $D^{\rm tSZ}_{3000} = 3.3\pm1.4\,\mu{\rm K}^2$ and $D^{\rm kSZ}_{3000} < 8.6\,\mu{\rm K}^2$ (95\% CL) at 150\,GHz. 
The SPT and ACT power spectra and resulting constraints on SZ power are consistent within the reported
uncertainties. 
On much larger angular scales ($\ell \le 2000$), the \planck{} collaboration has detected the tSZ power spectrum at high significance \citep{planck13-21}. 
The measured tSZ power at all scales is less than predicted by some models, suggesting either current astrophysical models over-predict the tSZ power or that $\sigma_8$ is significantly less than the value of $\sim$\,0.8-0.83 preferred by \wmap\, and \planck{} analyses of primary CMB anisotropy \citep{hinshaw13,planck13-16}.

The tSZ power has also been detected in higher-order moments of maps using data from ACT \citep{wilson12} and SPT \citep{crawford14} (hereafter C14).
Including an 11\% modeling uncertainty, C14 use the measurement of the bispectrum to predict the tSZ power to be $D^{\rm tSZ}_{3000} = 2.96\pm0.64\,\mu{\rm K}^2$ at 153\,GHz. 
The results are consistent with the direct power measurements. 

The same data used to measure the SZ power spectrum can also be used to measure the CIB power
at millimeter wavelengths and, if multi-frequency data are available, the CIB spectral behavior at these
wavelengths.
In previous work, the data lacked statistical power to test the need for separate spectral indices for 
the Poisson and clustered components.
R12 constrained the combined spectral index of both components between 150 and 220\,GHz to be $\alpha = 3.56 \pm 0.07$, 
 i.e.~the CIB flux scales with frequency as $F_\nu \propto \nu^{\alpha=3.56}$.  
When the two indices were allowed to vary independently, R12 measured corresponding indices of $\alpha_p = 3.45 \pm 0.11$ and $\alpha_c = 3.72 \pm 0.12$, a difference of 1.7$\sigma$.
\citet{dunkley13} also detected significant power attributed to clustered DSFGs in the ACT data and found a preferred combined DSFG spectral index of $3.7 \pm 0.1$, consistent with the R12 results. 
However, when allowing the spectral indices to vary independently, \citet{dunkley13} were unable to constrain the clustered component of the CIB. 
The spectral index depends on both the spectral energy distribution of individual DSFGs and their redshift distribution, and these quantities have also recently been determined statistically through cross correlations with ancillary catalogs \citep[]{viero13b}. 

When including a single-component model for correlation between tSZ power and CIB power, R12 found the normalized correlation coefficient at $\ell=3000$ to be $\xi = 0.18 \pm 0.12$.
The analysis of the ACT data presented in \citet{dunkley13} does not constrain this correlation; instead, a uniform prior is placed on the correlation.

R12 constrained the amplitude of the kSZ power at $\ell = 3000$ to be $D^{\rm kSZ}_{3000} < 2.8\,\mu{\rm K}^2$ at 95\% confidence in the absence of tSZ-CIB correlations.
When correlations were allowed, this constraint relaxed to $D^{\rm kSZ}_{3000} < 6.7\,\mu{\rm K}^2$ at 95\% confidence.
\citet{dunkley13} constrain the amplitude of the kSZ power to be $D^{\rm kSZ}_{3000} < 8.6\,\mu{\rm K}^2$ at 95\% confidence with a uniform prior of $\xi \in [0, 0.2]$. 
When this prior was relaxed to $\xi < 0.5$ in \citet{sievers13}, the kSZ constraint was weakened to $D^{\rm kSZ}_{3000} < 9.4\,\mu{\rm K}^2$ at 95\% confidence.

The kSZ power can be interpreted as a constraint on reionization. \citet[][hereafter Z12]{zahn12} used reionization simulations to show that when all of the kSZ power in the baseline model R12 is interpreted to be from reionization (a more conservative assumption in the case of this upper limit than attributing some of it to homogeneous kSZ), the upper limit on the duration of the period in which the universe goes from an average ionization of $\bar{x}_{e}$ =0.20 to 0.99  is $\Delta z \equiv z_{\xhii=0.20}-z_{\xhii=0.99} \le 4.4$ (95\% confidence).
Z12 report a yet more conservative limit of $\Delta z \le 7.9$ (95\% confidence) when allowing for tSZ-CIB correlation.

\subsection{This work}
This is the fourth SPT power spectrum analysis focused on CIB and secondary CMB anisotropies.  
Here, we improve upon the last release by R12 in two ways. 
First and most importantly, we use the full 2540\,deg$^2$ \sptsz{} survey, which covers three times (four times at 95\,GHz) more sky area than used by R12  with the concomitant reduction in power spectrum uncertainties. 
We also include twice as much data on 200\,deg$^2$ of the R12 region. 
These are the two SPT ``deep" fields which were first observed in 2008, and reobserved to the same noise level at 150\,GHz in 2010 and 2011. 
Second, we improve the analysis by more optimally weighting the data in two-dimensional Fourier space which reduces the power spectrum uncertainties by approximately 15\%.

The paper is organized as follows. 
The observations and power spectrum analysis are described in Section \ref{sec:dataanalysis}, and systematics checks are detailed in Section \ref{sec:jackknife}. 
The resulting power spectra are presented in Section \ref{sec:bandpowers}. 
The model to which the data are fit is presented in Section \ref{sec:model}. 
An overview of the cosmological constraints is presented in Section \ref{sec:results}, with more detailed investigations of the implications for the tSZ, the kSZ, the tSZ-CIB correlation, and CIB power spectra in Sections \ref{sec:tszinterp}, \ref{sec:kszinterp}, \ref{sec:tszcibinterp}, and \ref{sec:cibinterp} respectively. 
We conclude in Section \ref{sec:conclusions}.

\section{Data and analysis}
\label{sec:dataanalysis}

We present power spectra from 2540 square degrees of SPT data at 95, 150, and 220\,GHz.
Power spectra are estimated using a pseudo-${\it C}_\ell$ cross-spectrum method \citep{hivon02,polenta05,tristram05}. 
We calibrate the data by comparing the power spectrum to \emph{Planck} 1-year CMB power spectrum \citep{planck13-15}. 

\subsection{Data}
\label{subsec:data}

The 2540 \degsq\ of sky analyzed in this work were observed with the \sptsz{} camera on the SPT from 2008 to 2011. 
The full survey was divided into 19 contiguous sub-patches, referred to as fields, for observations.   
The specific field locations and extents can be found in Table 1 of \citet{story13}, hereafter S13. 
The \sptsz{} camera consists of 960 horn-coupled spiderweb bolometers with 
superconducting transition edge sensors and was installed on the 10-meter SPT from 2007-2011. 
The telescope, receiver, and detectors are discussed in more detail in \citet{ruhl04}, \citet{padin08}, \citet{shirokoff09}, and \citet{carlstrom11}. 

Seventeen of the 19 fields were observed to an approximate depth of 18 $\mu$K-arcmin at 150\,GHz. 
The noise levels at 95 and 220\,GHz vary significantly between 2008 and subsequent years as the focal plane was refurbished to maximize sensitivity to the tSZ effect before the 2009 observing season. 
This refurbishment added science-quality 95\,GHz detectors and removed half of the 220\,GHz detectors. 
As a result, the 2009-2011 data are comparatively shallower at 220\,GHz, but much deeper at 95\,GHz. 
The remaining two fields (200\,\sqdeg), first observed in 2008, were reobserved in 2010 and 2011 to acquire higher quality 95\,GHz data. 
Thus these two fields have $\sim$$\sqrt{2}$ lower noise at 150\,GHz. 
The approximate statistical weight of data from each year [2008, 2009, 2010, 2011] is [0, 0.18, 0.42, 0.40] at 95\,GHz, [0.08, 0.19, 0.40, 0.33] at 150\,GHz, and [0.20, 0.17, 0.39, 0.25] at 220\,GHz.
Details of the time-ordered data (TOD), filtering, and map-making can be found in S11.

\subsection{Beams and calibration}
\label{sec:beamcal}

The SPT beams are measured using a combination of bright point sources in each field, Venus, and Jupiter as described in S11. 
The main lobes of the SPT beam at 95, 150, and 220$\,$GHz are well-represented by 1.7$^\prime$, 1.2$^\prime$, and 1.0$^\prime$ FWHM Gaussians.

The observation-to-observation relative calibrations of the TOD are determined from measurements of a galactic H{\small \,II} region, RCW38, repeated several times per day.
The absolute calibration at each frequency is determined by comparing the \planck{} combined CMB power spectrum and the single-band SPT power spectra over the multipole range $\ell \in [670,1170]$ in $\ell$-bins that match the released \planck{} power spectrum. 
Modes are equally weighted within each bin. 
The power in this multipole range is dominated by primary CMB power at our observing frequencies.
This calibration method is model independent as it only requires that the primary CMB power in the SPT fields be statistically representative of the all-sky power.
We estimate the uncertainty in the SPT power calibration at [95, 150,  220] GHz to be [1.1\%, 1.2\%, 2.2\%].  
The correlated part of these uncertainties is 1.0\%.
The high correlation is unsurprising, as the uncertainty in the calibration is dominated by SPT sample variance, which is nearly 100\% correlated between the three SPT bands.
The beam and calibration uncertainties are included in the power spectrum covariance matrix as described in Section \ref{sec:beamunc}.

\subsection{Power spectrum estimation}
\label{sec:bandpowerest} 

To estimate the power spectrum, we use a pseudo-$C_\ell$ method \citep{hivon02}.  
 Pseudo-$C_\ell$ methods estimate the power spectrum from the spherical harmonic transform of the map, after correcting for effects such as TOD filtering, beams, and finite sky coverage. 
 We use  the MASTER algorithm \citep{hivon02} to correct for beam, apodization, and filtering effects. 
In a departure from the MASTER algorithm, however, we calculate cross-spectra \citep{polenta05, tristram05} to eliminate noise bias in the spectrum. 
The SPT observation strategy is well-suited to cross-spectra as it produces approximately 100 complete, independent maps for each field. 
The analysis is performed in the flat-sky approximation and Fourier transforms are used instead of spherical harmonic transforms. 
We report the power spectrum in terms of $\mathcal{D}_\ell$, where
\begin{equation}
\mathcal{D}_\ell=\frac{\ell\left(\ell+1\right)}{2\pi} C_\ell\;.
\end{equation}

Details of the power spectrum estimator have been presented in previous SPT papers, most relevantly L10, \citet{keisler11} (hereafter K11), and R12. 
In the following sections, we provide an overview on the method and note differences between the current analysis and the method used by R12.

\subsubsection{Cross spectra}
\label{sec:xspec}
We apply a window to the map which smoothly goes to zero at the map edges and masks point sources detected above 5$\sigma$ at 150 GHz. 
The individual point source masks are a 2 arcmin radius disc for sources detected between 5$\sigma$ and 50$\sigma$, with the disc size increased to 5 arcmin for sources detected above 50$\sigma$. 
Outside of this radius, a Gaussian taper with $\sigma_{taper}=5$ arcmin is applied.  

We Fourier transform the windowed maps, and then take a weighted average of the two-dimensional power
spectrum within an $\ell$-bin $b$,
\begin{equation}
\label{eqn:ddef}
 \widehat{D}^{\nu_i x \nu_j, AB}_b\equiv \left< \frac{\ell(\ell+1)}{2\pi}H_\ell \mathrm{Re}\left[\tilde{m}^{\nu_i,A}_{\ell}\tilde{m}^{\nu_j,B*}_{\ell}\right] \right>_{\ell \in b}, 
\end{equation} 
where $H_\ell$ is a two-dimensional weight array described below and $\tilde{m}^{\nu_i, A}$ is the Fourier transformed map. 
Here, $\nu_i, \nu_j$ are the observation frequencies (e.g., 150\,GHz) and $A, B$ are the observation indices. 
We average all cross-spectra $\widehat{D}^{AB}_b$ for $A \neq B$ to calculate  binned power spectrum $\widehat{D}_b$ for each field. 
We refer to the power within a bin $b$ as defined by equation \ref{eqn:ddef} as a ``bandpower."

The maps have anisotropic noise due to the observation strategy and TOD filtering. 
At low $\ell$, modes perpendicular to the scan direction ($\ell_x=0$) are noisier than those parallel to the scan direction ($\ell_y = 0$). 
R12 addressed this by simply masking modes at $\ell_x < 1200$. 
In this work, we instead follow the weighting procedure used in K11.
We construct a two-dimensional weight array $H_\ell$, defined as 
\begin{equation}
\label{eqn:2dweight}
H_{\ell} \propto (C_\ell^{\mathrm{th}} + N_{\ell})^{-2} \,,
\end{equation}
 where $C_\ell^{\mathrm{th}}$ is the theoretical power spectrum used in simulations described in section \ref{sec:sims}, and $N_{\ell}$ is the two-dimensional, calibrated, beam-deconvolved noise power. 
 $N_\ell$ is calculated from difference maps between scans going in opposite directions. 
 In most cases, this is the difference between left-going and right-going scans, however, for observations which used  elevation scans, this is the difference between upward-going and downward going scans. 
The weight array is smoothed with a Gaussian kernel of width $\sigma_{\ell}=1000$ to reduce the scatter in the noise power, and is normalized such that $\sum_{\ell \in b}H_{\ell}=1$ for each bin $b$.
A separate weight array is calculated for each observation field. 
Application of the weight array improves the signal-to-noise of the power spectrum by approximately 15\%. 

\subsubsection{Simulations}
\label{sec:sims}

Noise-free simulations are used to estimate the sample variance and transfer function from map-making. 
For each field, we create 100 sky realizations that are smoothed by the appropriate beam for each frequency.
We then simulate observations of these sky realizations by sampling each realization using the SPT pointing information. 
These simulated TOD are  filtered and processed into maps identically to the real data.  

Each simulated sky is the sum of  Gaussian realizations of the best-fit lensed WMAP7+S13 $\Lambda$CDM primary CMB model, a tSZ model, a kSZ model, and point source contributions. 
The simulation inputs have been updated to be consistent with the R12 bandpowers, which have point sources masked at the same significance level as this work, corresponding to a flux cut above $\sim6.4$ mJy at 150\, GHz.
The kSZ power spectrum is based on the \citet{sehgal10} simulations with an amplitude of $2.0\, \mu{\rm K}^2$ at $\ell=3000$. 
The tSZ power spectrum is calculated from the \citet{shaw10} simulations and has an amplitude of $4.4\, \mu{\rm K}^2$ at $\ell=3000$ in the  150\,GHz band. 
There are three point source components in the simulations: terms from both spatially clustered and Poisson-distributed dusty star-forming galaxies (DSFGs), and a term for Poisson-distributed radio sources.
The DSFG Poisson term has an amplitude of $D_{3000}^{p} = 7.7 \,\mu{\rm K}^2$ and the clustered DSFG component is modeled by a $D_\ell \propto \ell^{0.8}$ term normalized to $D_{3000}^{c} = 5.9 \,\mu{\rm K}^2$, both at 150 GHz.
 A spectral index of 3.6 is assumed for both DSFG components. 
 Based on the \citet{dezotti05} model with the 6.4 mJy flux cut for SPT sources, the radio power is set to $D_{3000}^{r} = 1.28\, \mu{\rm K}^2$ at $150\,$GHz with an assumed spectral index of $\alpha_r=-0.6$
 These simulations neglect non-Gaussianity in the tSZ, kSZ, and radio source contributions and thus slightly underestimate the sample variance. 
\citet{millea12} argue this is negligible for SPT since the bandpower uncertainties are dominated by instrumental noise on the relevant angular scales. 

\subsubsection{Covariance estimation and conditioning}
\label{sec:cov}

The bandpower covariance matrix accounts for sample variance and instrumental noise variance.  
As in R12, the sample variance is estimated from the signal-only  simulations (see Section \ref{sec:sims}),  and the noise variance is calculated from the distribution of the cross-spectrum bandpowers $D^{\nu_i\times\nu_j,AB}_b$ between observations A and B, and frequencies $\nu_i$ and $\nu_j$. 
At $\ell > 5000$, where the signal is strongly noise dominated, we ensure the covariance matrix remains positive definite by using only the first term in eqn.~A15 in L10. 
This term scales with the variance of the cross spectra, and is an accurate estimate of the covariance in the noise-dominated regime.
There is statistical uncertainty on elements of the estimated bandpower covariance matrix that can degrade parameter constraints \citep[see, e.g.,][]{dodelson13}. 
We ``condition'' the covariance matrix to reduce this uncertainty following the prescription in R12. 
We find in simulations that this conditioning significantly reduces the degradation in parameter constraints. 

The covariance matrix is a set of thirty-six square blocks, with the six on-diagonal blocks corresponding to the covariances of 95 x 95, 95 x 150, ... 220 x 220 GHz spectra. 
The covariance matrix includes an estimate of the signal, and, if both bandpowers share a common map, noise variance.
In the estimated covariances of off-diagonal blocks for which the instrumental noise term is included, e.g., the ($95\times 150, 150\times 220$) covariance, the uncertainty in the covariance estimate can be large compared to the true covariance.
Therefore we calculate the elements of these off-diagonal blocks analytically from the diagonal elements of the on-diagonal blocks using the equations in Appendix A of L10. 

\subsubsection{Field weighting}
\label{sec:fieldweighting}

We calculate a weight, $w^{i}$, for each field and frequency cross spectrum in order to combine the individual-field bandpowers. 
Unlike R12 who use an $\ell$-dependent weighting scheme,  we set $w^{i}$ for each frequency combination equal to the average of the inverse of the diagonal of the covariance matrix over the bins $2500<\ell<3500$.
Note that this covariance matrix includes noise and sample variance from simulations, but no beam or calibration errors.
We expect and observe that the optimal weights vary slowly with angular scale, and these angular scales are chosen to maximize the sensitivity to the SZ signals.

The combined bandpowers are calculated from the individual-field bandpowers according to:
\begin{equation}
D_b = \sum_{i}D_{b}^{i}w^{i}
\end{equation}
and the covariance matrix similarly:
\begin{equation}
\textbf{C}_{bb^\prime} = \sum_{i}w^{i}\textbf{C}_{bb^\prime}^{i}w^{i}.
\end{equation}
Note that the sum of the weights $\sum_{i}w^{i} = 1$.

\subsubsection{Beam and calibration uncertainties}
\label{sec:beamunc}

For the cosmological analysis, we fold beam and calibration uncertainties into the bandpower covariance matrix. 
This follows the treatment of beam uncertainties laid out by K11 and used by R12. 
Calibration uncertainty mimics a constant fractional beam uncertainty. 

As in K11, we estimate a beam-calibration correlation matrix for each year:
\begin{equation}
\pmb{\rho}^{\rm beam}_{bb^\prime} = \left(\frac{\delta D_b}{D_b}\right) \left(\frac{\delta D_{b^\prime}}{D_{b^\prime}}\right)
\end{equation}
where 
\begin{equation}
\frac{\delta D_b}{D_b} = 1-\left(1+\frac{\delta B_b}{B_b}\right)^{-2},
\end{equation}
and $B_b$ is the Fourier transform of the map of the beam averaged across bin $b$.

We then combine the correlation matrix for each year according to the yearly weights, which is the sum of the fraction of the weights from each field in a given year at a given frequency.
We add to this the calibration correlation matrix defined by $\pmb{\rho}^{\rm cal} = \sigma_{\rm cal}^2$, where $\sigma_{\rm cal}$ is the fractional calibration uncertainty. 
The correlation matrix is transformed into a covariance matrix by multiplying by the measured bandpowers:

\begin{equation}
\textbf{C}^{\rm beam+cal}_{bb^\prime} = \pmb{\rho}^{\rm beam+cal}_{bb^\prime}D_{b}D_{b^\prime}.
\end{equation}

We  add this beam and calibration covariance to the bandpower covariance matrix which contains the effects of sample variance, and instrumental noise. 
The final covariance matrix includes beam, calibration, sample variance and instrumental noise. 

\section{Null tests}
\label{sec:jackknife}

We use null tests to search for systematic errors in the data. 
In each null test, we divide the set of maps into two halves and difference the halves to remove true astrophysical signals. 
We then apply the cross spectrum analysis described in the last section to the differenced maps, which should result in a spectrum consistent with zero. 

\begin{figure}[htb]\centering
\includegraphics[width=0.45\textwidth,clip,trim =  1.80cm  12.06cm  11cm  5.34cm]{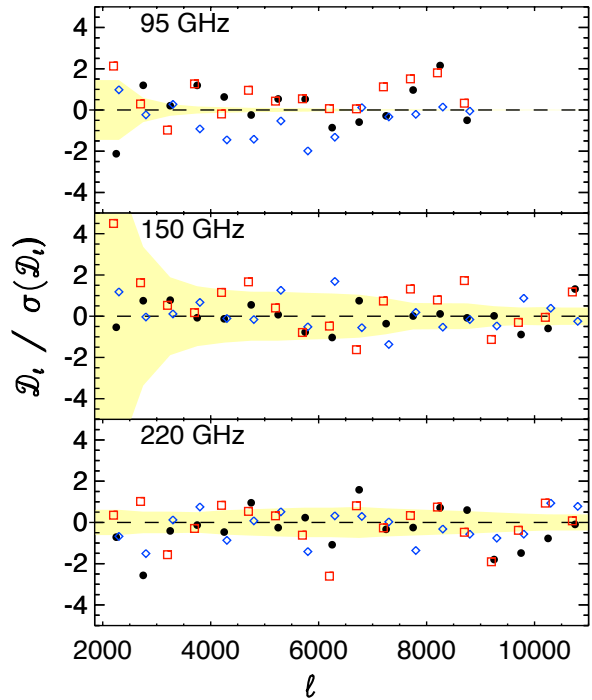}
  \caption[Null tests]{
  The results of the null tests applied to the SPT data, shown by the significance of each bin relative to zero power, where $\sigma(D_{\ell})$ is the noise-only error. 
  Thus, we expect the null-test points to have a Gaussian scatter with a size of 1 about zero.
  The yellow shaded region is for reference only and shows the sample variance uncertainty in units of $\sigma_\mathrm{samp}/\sigma(D_{\ell})$ for the undifferenced bandpowers. 
Any point within the yellow region has a value smaller than sample variance uncertainty, and hence does not bias the results of parameter fits, even if it is several $\sigma$ from zero.
  \textbf{Red squares} are the ``scan direction'' null spectra, \textbf{black circles} are the ``azimuthal range'' null spectra, and \textbf{blue diamonds} are the ``time'' null spectra. 
  The bandpowers for each null test are slightly offset in $\ell$ for clarity. 
   \textbf{Top panel}: 95 GHz null spectra.  \textbf{Second panel}: 150 GHz null spectra. 
  \textbf{Third panel}: 220 GHz null spectra.  
  \label{fig:jackknife}
  }
\end{figure}

In the limit of perfect knowledge of the instrument, any remaining signals should not be of astrophysical origin; the residual astrophysical signal in simulated null spectra is much smaller than the uncertainties. 
The various ways in which the data are split in half are chosen to maximize the sensitivity to likely sources of systematic errors. 
We compare the resulting spectra to the expectation value of zero, with a significant deviation from zero implying the presence of a systematic error. 
We look for systematic effects using the following data splits:

\begin{itemize}

\item Scan direction: We difference data from left-going and right-going scans (or in a minority of cases, upward and downward-going). 
This tests for effects sensitive to telescope velocity or acceleration.
Examples include errors in the estimated detector time constants or microphonic pickup.

\item Azimuthal Range: We difference the data depending on the observation azimuth to maximize
sensitivity to emission from the ground and structures around the telescope.
We group the data according to the magnitude of observed ground emission on several degree scales, as detailed by S11.

\item Time: We difference the data based on observation time.
We compare data from the first half of the observations of a field to  the second half.
This tests the temporal stability of the instrument, including calibration, pointing, detector responsivity and linearity, and any other long time-scale effect.

\end{itemize}
We find the results of each test to be consistent with zero. 
The results are plotted in Figure  \ref{fig:jackknife}. 

Combining all three frequencies, the probabilities to exceed (PTE) for the  ``scan direction", ``azimuthal range", and ``time" null tests are 0.02, 0.79, and 0.96 respectively. 
At 95, 150, and 220 GHz, the combined PTEs for the three tests combined are 0.43, 0.37, and 0.68. 
Table \ref{tab:PTE} shows the PTEs for the nine individual null tests.
We conclude that there is no evidence for systematic biases. 

\begin{table}[htb]
\begin{center}
\caption{\label{tab:PTE} PTE for Null Tests}
\small
\begin{tabular}{l || c c c | c}
\hline\hline
\rule[-2mm]{0mm}{6mm}
  & 95 GHz & 150 GHz & 220 GHz & Combined \\
\hline
\hline
Scan dir. &0.34 &0.003 & 0.44 & 0.02\\
AZ range &0.35 &0.98 &0.37 & 0.79 \\
Time &0.59 &0.92 &0.89 & 0.96 \\
\hline
Combined & 0.43 & 0.37 & 0.68 & \\
\hline
\hline
\end{tabular}
\tablecomments{ 
PTEs for the 9 individual null tests, and combined values for each frequency and type of test. 
The low PTE for the 150 GHz scan direction test is due to a 4\,$\sigma$ deviation in the lowest $\ell$-bin. 
However, this residual power is well below the sample variance uncertainty in that bin (see Figure \ref{fig:jackknife}),  and hence does not bias the parameter fits.
} \normalsize
\end{center}
\end{table}

\section{Bandpowers}
\label{sec:bandpowers}

\begin{figure*}[t]\centering
\includegraphics[width=0.95\textwidth,clip,trim =  1.59cm  12.02cm  6.83cm  6.54cm]{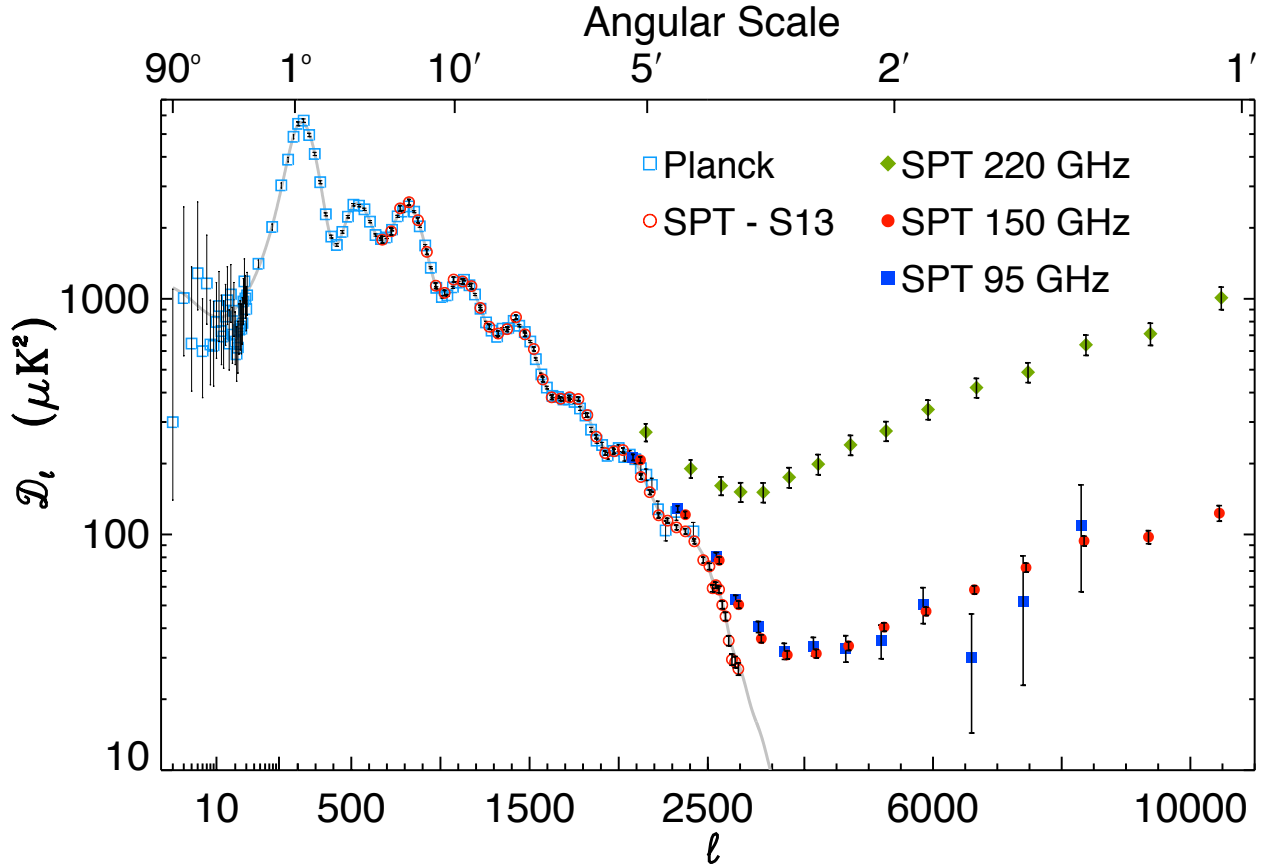}
  \caption[]{
  The SPT and \planck{} bandpowers. 
  The \planck{} and S13 bandpowers (open squares and circles)  are primary CMB only, and agree well on all angular scales.
  The grey line is the lensed $\Lambda$CDM CMB theory spectrum. 
  We also show bandpowers at 95, 150, and 220\,GHz (filled squares, circles and diamonds) measured with the SPT in this work. 
 On large scales, the primary CMB anisotropy is dominant at all frequencies. 
  On smaller scales, contributions from the CIB, radio sources, and secondary CMB anisotropies (tSZ and kSZ) dominate the observed power.
  The observed differences between frequency bands are due to these other sources of power. 
  The CIB dominates the power spectrum at small scales at 150 and 220\,GHz; radio galaxies are more important at 95\,GHz. 
    }
  \label{fig:bandpowersall}
\end{figure*}

The analysis described in Section \ref{sec:bandpowerest} is applied to the 2540 \sqdeg\ of sky observed in the \sptsz{} survey. 
The resulting bandpowers are shown in Figure \ref{fig:bandpowersall} and tabulated in Table \ref{tab:bandpowers}. 
All bandpowers presented have point sources detected at $>5\sigma$ ($\sim 6.4$ mJy) at 150 GHz masked.
The bandpowers, covariance matrix, and window functions are available for download on the SPT\footnote{\label{footnote:spt}http://pole.uchicago.edu/public/data/george14/} 
and LAMBDA\footnote{\label{footnote:lambda}http://lambda.gsfc.nasa.gov/product/spt/spt\_prod\_table.cfm} websites. 
The websites also have bandpowers with galaxy clusters masked (see Section \ref{sec:clustermask}). 

\begin{figure*}[t]\centering
\includegraphics[trim=1.8cm 1.5cm 9cm 8cm,clip,width=0.9\textwidth]{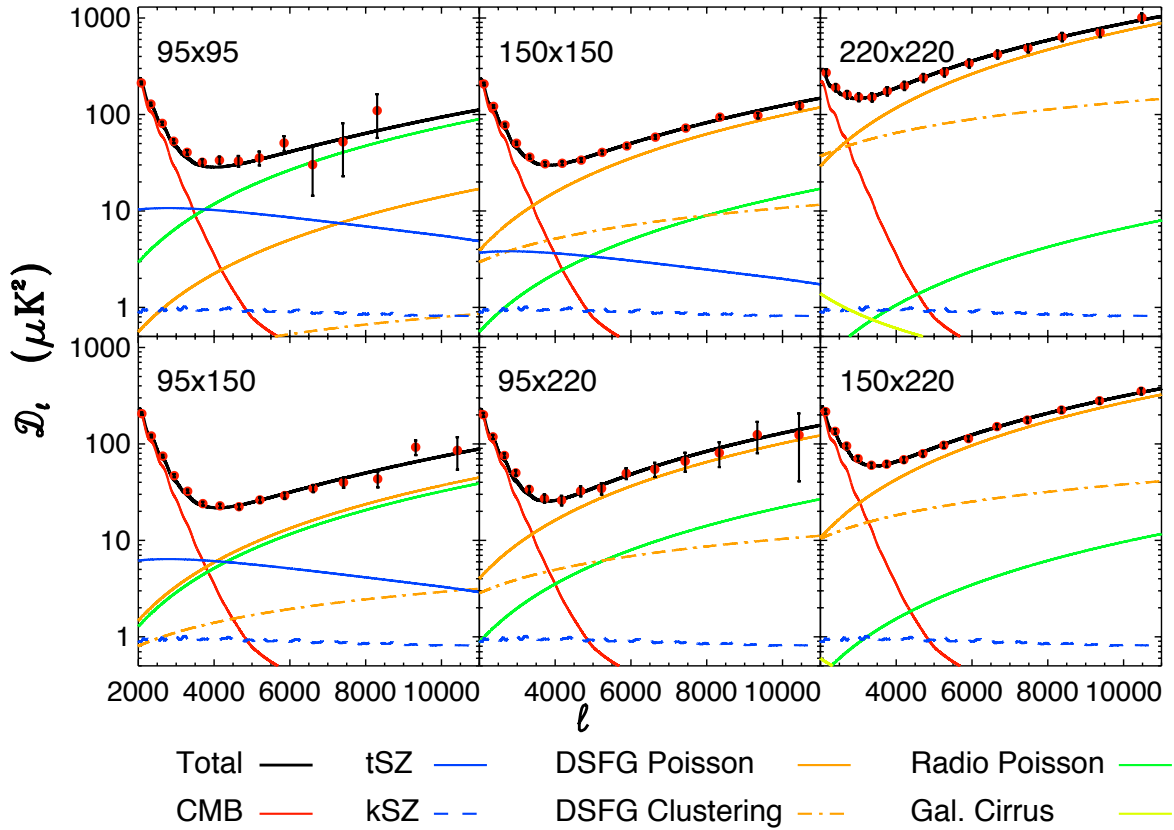}
\caption[]{ 
The six auto- and cross-spectra measured with the 95, 150, and 220\,GHz SPT data. 
  Overplotted on the bandpowers is the best-fit model for the fiducial set of model parameters. 
  The bandpowers have not been corrected by the best-fit calibration or beam uncertainties. 
  }
  \label{fig:bandpowerbestfit}
\end{figure*}

At the SPT observing frequencies, the power spectrum at $\ell < 3000$ is dominated by the primary CMB anisotropy. 
At approximately $\ell = 3000$, the cosmic infrared background (CIB) anisotropy becomes dominant, which includes both clustered and  Poisson components and asymptotes to an $\ell^2$ form at high $\ell$. 
In the 95\,GHz band, power from Poisson distributed radio sources is dominant for large $\ell$.
We detect the tSZ effect, constrain a  correlation between the tSZ and CIB powers, and set upper limits on the amplitude of the kSZ signal. 
Figure \ref{fig:bandpowerbestfit} shows the bandpowers in our 6 frequency combinations, including the best-fit model components.

\begin{table*}[ht!]
\begin{center}
\caption{\label{tab:bandpowers} Bandpowers}
\small
\begin{tabular}{cc|cc|cc|cc}
\hline\hline
\rule[-2mm]{0mm}{6mm}
& &\multicolumn{2}{c}{$95\,$GHz} &\multicolumn{2}{c}{$150\,$GHz} & \multicolumn{2}{c}{$220\,$GHz} \\
$\ell$ range&$\ell_{\rm eff}$ &$\hat{D}$ ($\mu{\rm K}^2$)& $\sigma$ ($\mu{\rm K}^2$) &$\hat{D}$ ($\mu{\rm K}^2$)& $\sigma$ ($\mu{\rm K}^2$)&$\hat{D}$ ($\mu{\rm K}^2$)& $\sigma$ ($\mu{\rm K}^2$) \\
\hline

 2001 -  2200 &  2106 & 212.4 &   6.3 & 207.1 &   6.3 &  271.6 &  23.4 \\ 
 2201 -  2500 &  2357 & 128.3 &   4.0 & 121.3 &   3.8 &  190.3 &  16.7 \\ 
 2501 -  2800 &  2657 &  80.9 &   2.9 &  77.6 &   2.6 &  161.0 &  14.5 \\ 
 2801 -  3100 &  2958 &  52.8 &   2.4 &  50.4 &   1.8 &  151.5 &  14.3 \\ 
 3101 -  3500 &  3308 &  40.5 &   2.2 &  36.2 &   1.4 &  151.0 &  14.4 \\ 
 3501 -  3900 &  3709 &  32.0 &   2.5 &  30.8 &   1.2 &  174.9 &  17.1 \\ 
 3901 -  4400 &  4159 &  33.5 &   3.0 &  31.2 &   1.3 &  198.8 &  19.5 \\ 
 4401 -  4900 &  4660 &  33.0 &   4.2 &  33.7 &   1.4 &  239.9 &  23.3 \\ 
 4901 -  5500 &  5210 &  35.5 &   5.8 &  40.5 &   1.7 &  274.9 &  26.2 \\ 
 5501 -  6200 &  5861 &  50.7 &   8.8 &  47.2 &   1.9 &  339.1 &  32.0 \\ 
 6201 -  7000 &  6612 &  30.2 &  15.8 &  58.3 &   2.3 &  419.8 &  39.6 \\ 
 7001 -  7800 &  7412 &  52.0 &  29.1 &  72.3 &   3.2 &  488.0 &  47.2 \\ 
 7801 -  8800 &  8313 & 109.7 &  52.6 &  93.9 &   4.5 &  638.1 &  63.2 \\ 
 8801 -  9800 &  9313 &   - &   - &  97.6 &   6.2 &  710.5 &  76.5 \\ 
 9801 - 11000 & 10413 &   - &   - & 123.2 &   9.4 & 1010.4 & 111.1 \\

\hline
&&\multicolumn{6}{c}{}\\
& & \multicolumn{2}{c}{$95\times150\,$GHz} & \multicolumn{2}{c}{$95\times220\,$GHz} & \multicolumn{2}{c}{$150\times220\,$GHz} \\
\hline

 2001 -  2200 &  2106 & 205.8 &   5.6 & 200.3 &   9.6 &  216.0 &  11.2 \\ 
 2201 -  2500 &  2357 & 120.8 &   3.4 & 118.3 &   5.8 &  134.9 &   7.2 \\ 
 2501 -  2800 &  2657 &  74.7 &   2.3 &  75.9 &   4.0 &   95.2 &   5.2 \\ 
 2801 -  3100 &  2958 &  46.9 &   1.6 &  50.1 &   3.1 &   70.4 &   4.1 \\ 
 3101 -  3500 &  3308 &  32.3 &   1.2 &  33.8 &   2.5 &   60.1 &   3.6 \\ 
 3501 -  3900 &  3709 &  24.0 &   1.0 &  27.2 &   2.7 &   61.7 &   3.8 \\ 
 3901 -  4400 &  4159 &  22.8 &   1.1 &  26.0 &   3.0 &   68.6 &   4.3 \\ 
 4401 -  4900 &  4660 &  22.3 &   1.3 &  32.5 &   3.9 &   79.2 &   5.0 \\ 
 4901 -  5500 &  5210 &  26.3 &   1.5 &  34.5 &   4.6 &   96.8 &   5.9 \\ 
 5501 -  6200 &  5861 &  29.1 &   2.0 &  49.6 &   6.4 &  113.5 &   6.8 \\ 
 6201 -  7000 &  6612 &  34.6 &   2.9 &  54.5 &   9.2 &  151.2 &   8.9 \\ 
 7001 -  7800 &  7412 &  39.9 &   4.8 &  66.2 &  14.8 &  176.6 &  10.7 \\ 
 7801 -  8800 &  8313 &  43.3 &   7.9 &  80.8 &  23.3 &  224.4 &  14.1 \\ 
 8801 -  9800 &  9313 &  92.5 &  16.0 & 124.8 &  44.6 &  280.0 &  19.2 \\ 
 9801 - 11000 & 10413 &  85.8 &  31.8 & 123.9 &  83.0 &  352.4 &  26.0 \\

\hline
\end{tabular}
\tablecomments{Angular multipole range, weighted multipole value $\ell_{\rm eff}$, bandpower $\hat{D}$, 
and bandpower uncertainty $\sigma$ for the six auto and cross-spectra of the $95\,$GHz, $150\,$GHz, and $220\,$GHz maps with point sources detected at $>5 \sigma$ ($\sim 6.4\,$mJy) at 150 GHz masked at all frequencies.
The uncertainties in the table are calculated from the diagonal elements of the covariance matrix, which includes sample variance, beam errors, calibration errors (see Section \ref{sec:beamcal}), and instrumental noise. 
Due to the larger beam size at 95\,GHz (and resulting lower S/N at high $\ell$), the 95x95\,GHz bandpowers are limited to $\ell < 8800$. 
}
\normalsize
\end{center}
\end{table*}

\section{Cosmological modeling}
\label{sec:model}

We fit the SPT bandpowers to a model that includes lensed primary CMB anisotropy, tSZ and kSZ effects, radio galaxies, the dusty galaxies that make up the CIB, and correlations between the tSZ signal and CIB power.
DSFGs are the source of the CIB and are the most important foreground, especially at high frequencies. 
We model the power spectrum of DSFGs as the sum of unclustered (Poisson) and clustered components. 
We also include a Poisson contribution from radio galaxies that is most important at lower frequencies.  
Finally, we include a term for the galactic cirrus. 
This final component is negligible, as we have chosen to observe fields with low galactic emission.

Parameter constraints are calculated using the November 2013 version of the publicly available {\textsc CosmoMC}\footnote{http://cosmologist.info/cosmomc} package \citep{lewis02b}.
We add two modules to the standard {\textsc CosmoMC} for fitting the high-$\ell$ data: one to model the foregrounds and secondary anisotropies and one to calculate the SPT  likelihood function. 
The code is a modified and expanded version of the modules discussed in \citet{millea12} and used by S11 and R12. 
These new modules and instructions for compiling them are available at the SPT website.$^{\ref{footnote:spt}}$

The cosmological constraints presented here include measurements of the primary CMB anisotropy, baryon acoustic oscillations (BAO), and Hubble constant (H$_0$).
For the primary CMB, we include the \planck\ 2013 data release \citep{planck13-15}. 
Due to the different frequencies and source cuts, we use different foreground models for the Planck and SPT data. 
The Planck likelihood uses the published Planck foreground model; the SPT-SZ likelihood uses the foreground model described above. 
We share the three variables between the \planck\ and SPT-SZ likelihoods that describe similar quantities in each dataset: the tSZ power, kSZ power and the tSZ-CIB correlation, while otherwise treating the models and their foreground parameters as independent.
We do not include the low-$\ell$ SPT bandpowers of S12 for two reasons. 
At the small angular scales where the error bars are comparable or better than \planck{}, they would match the bandpowers used in this work. 
At lower $\ell$s, \planck\ completely dominates regardless of whether S12 is used. 
Thus, including the S12 band powers has no effect on the derived secondary parameters.
The H$_0$ measurement used here is from the Hubble Space Telescope \citep[\emph{HST},][]{riess11}. 
We include measurements of the BAO feature from SDSS \citep{anderson12, padmanabhan12} and 6dF data \citep{beutler11}.  

\subsection{Cosmic microwave background anisotropy} 
\label{sec:primarycmb}

We predict the primary CMB temperature anisotropy within the standard, six-parameter, spatially flat, lensed $\Lambda$CDM cosmological model. 
For faster calculations, we use PICO\footnote{https://sites.google.com/a/ucdavis.edu/pico/} \citep{fendt07a,fendt07b} instead of CAMB \citep{lewis00} to calculate the primary CMB anisotropy. 

\subsection{Thermal Sunyaev-Zel'dovich anisotropy}
\label{sec:tsz}

We consider three models for the tSZ power spectrum.  
We adopt the baseline model from \citet{shaw10} as the baseline model in this work. 
Models by \citet{sehgal10} and \citet{bhattacharya12} are used as alternatives. 
The three models are similar in their approach (combining N-body simulation results
with semi-analytical models for gas physics) and
predict different amplitudes, but essentially the same angular shape, for the tSZ power spectrum. 
We briefly describe each model below.

The \citet{sehgal10} model (which we will refer to hereafter as the Sehgal model) is the earliest of the three models, and the only one produced before the L10 tSZ power results were published.
The Sehgal model combines the semi-analytic model for the intra-cluster medium (ICM) of \citet{bode09} with a cosmological N-body simulation to produce simulated tSZ and kSZ maps. 
The principal difference between this model and the later Shaw and Bhattacharya models is the lack of non-thermal pressure support from e.g., merger-induced shocks. 
Non-thermal pressure support reduces the tSZ signal in the outskirts of clusters. 
For assumed cosmological parameters of ($\Omega_b$, $\Omega_m$, $\Omega_\Lambda$, $h$, $n_s$, $\sigma_8$) = (0.044, 0.264, 0.736, 0.71, 0.96, 0.80), the Sehgal model predicts $D^{\rm tSZ}_{3000}=9.5\, \mu{\rm K}^2$ at 143\,GHz. 

In response to the results of L10, which showed very low tSZ power compared to the predictions of the Sehgal model, \citet{shaw10} investigated the impact of cluster astrophysics on the tSZ power spectrum, using a modified version of the \citet{bode09} model for gas physics. 
These modifications included adding components that suppress the tSZ power such as AGN feedback and non-thermal pressure support \citep[see also][]{battaglia12, trac11}.  
For the fiducial cosmological parameters given above, the baseline model from \citet[][hereafter the Shaw model]{shaw10} predicts $D^{\rm tSZ}_{3000}=5.5 \,\mu{\rm K}^2$ at $\ell =3000$ and 143\,GHz. 
We use this tSZ template in all Markov chain Monte Carlo (MCMC) chains where another model is not explicitly specified.

\citet{shaw10} also describe how the tSZ power scales with cosmological
parameters.
Around the fiducial cosmological model listed above, the tSZ power scales as
\begin{equation}
D^{\mathrm{tSZ}}  \propto \left(\frac{h}{0.71}\right)^{1.7} \left(\frac{\sigma_8}{0.80}\right)^{8.3} \left(\frac{\Omega_b}{0.044}\right)^{2.8}.
\label{eq:tsz_scaling}
\end{equation}
In chains in which we measure the fit quality to fixed tSZ templates, this equation is used to rescale the tSZ model templates as a function of cosmological parameters at each chain step, as described in S11.
We assume all tSZ models follow this cosmological scaling. 

The \citet{bhattacharya12} (Bhattacharya) model is identical to the Shaw model, except with different fiducial values for the astrophysical parameters. 
In particular, the energy feedback is lower and the non-thermal pressure support is higher. 
Taken together, this reduces the predicted tSZ power with the fiducial cosmological parameter values to $D^{\rm tSZ}_{3000}=5.0\, \mu{\rm K}^2$ at $143\,$GHz.

\subsection{Kinematic Sunyaev-Zel'dovich anisotropy}
\label{sec:ksz}

The kSZ power spectrum is naturally split into two model components. 
The ``homogeneous kSZ" component arises after reionization from the correlation of density fluctuations with the velocity field in the completely ionized universe.  
The second ``patchy kSZ" component comes from the motion of partially ionized gas during the epoch of reionization. 
The patchy kSZ power depends primarily on the duration of the reionization epoch.  
If reionization were instantaneous or isotropic, the patchy kSZ signal would be zero, and the only detectable kSZ signal would be the homogeneous kSZ signal from the post-reionization era.

We discuss our models for each component next.
The baseline template used to fit the kSZ signal in the chains is an equally weighted sum of the homogeneous and patchy kSZ templates. 

\subsubsection{Homogeneous Kinematic Sunyaev-Zel'dovich anisotropy}
\label{sec:hksz}

We rely upon the analytic model  of \citet{shaw12} that includes cooling + star formation (CSF) for the homogeneous kSZ  power spectrum. 
The model is calibrated to a hydrodynamical simulation, which includes metallicity-dependent radiative cooling and star-formation. 
The kSZ power spectrum was calculated through measurements of the power spectrum of gas fluctuations over a range of redshifts in each simulation. 
For the assumed fiducial cosmology (detailed in Section \ref{sec:tsz}), the CSF simulation predicts 1.6 \uksq\ at $\ell = 3000$. 
In this model, reionization occurs instantaneously at $\zend = 8$.
\citet{shaw12} argue that the CSF model is a robust lower limit on the amplitude of the hkSZ amplitude, and find that the total theoretical uncertainty on the amplitude of the hkSZ power at $\ell = 3000$ is 30\%.
This uncertainty is due to a combination of uncertainties on astrophysical processes, cosmological parameters, and helium reionization.

The approximate scaling of the homogeneous kSZ signal with cosmological parameters around the fiducial cosmology given by \cite{shaw12} is
\begin{eqnarray}
D^{\rm kSZ}  &\propto& \left(\frac{h}{0.71}\right)^{1.7} \left(\frac{\sigma_8}{0.80}\right)^{4.7} \left(\frac{\Omega_b}{0.044}\right)^{2.1}\times\nonumber\\
&& \left(\frac{\Omega_m}{0.264}\right)^{-0.4} \left(\frac{n_s}{0.96}\right)^{-0.2}.
\label{eq:ksz_scaling}
\end{eqnarray}
We use this scaling to rescale the kSZ CSF templates as a function of cosmological parameters in chains labeled ``Fixed CSF''. 

\subsubsection{Kinematic Sunyaev-Zel'dovich Anisotropy from patchy reionization}
\label{sec:pksz}

The patchy kSZ signal is modeled using the technique presented by Z12, which we briefly review here. 
The signal is determined from a simulation in which the matter over-density is calculated with a linear model.
The halo collapse fraction is set as a function of the over-density. 
The number of ionizing photons in a region is then estimated based on the number of collapsed halos above the cooling threshold mass. 
If there are sufficient ionizing photons given the number of hydrogen atoms within the halo, the halo is labeled ionized.
If not, this procedure is repeated at smaller and smaller scales until the simulation resolution is reached, which results in an ionization field. 
The ionized regions are then correlated with the density and velocity fields to determine the kSZ power spectrum.

The kSZ power spectrum resulting from these simulations is dependent upon two parameters: the midpoint of reionization $\zmid = z_{\bar{x}_{e}=0.50}$, where $\bar{x}_{e}$ is the average ionization fraction, and the duration of reionization, $\Delta z = z_{\bar{x}_{e}=0.20} - z_{\bar{x}_{e}=0.99}$. 
The choice of a relatively large ionization fraction of $\bar{x}_{e}=0.20$ for the beginning of reionization is due to the fact that lower ionized fractions are largely made up of ionized regions too small to be probed by the SPT data, hence we do not attempt to constrain the reionization state preceding this level of ionization.
The reionization model used in this analysis does not include self-regulation, and is therefore roughly symmetric about the midpoint of reionization.
Additionally, at least one study \citep{park13} has found that self-regulation can result in a relationship between
ionization history and patchy kSZ that is more complicated than the clean dependence on \zmid\ and 
$\Delta z$ found
by Z12 and other authors \citep[e.g.,][]{battaglia13,mesinger12}. For the purposes of this work, we 
use the Z12 model with no modifications and consider the effects of self-regulation as a potential caveat to our conclusions
about $\Delta z$.

Z12 also demonstrate that the shape of the patchy kSZ power spectrum from the simulations is quite robust to changes in duration and timing of reionization, implying that a single $\ell$-space template can be used for fitting the data. 
This conclusion is supported by the relatively small observed changes in the kSZ constraints from the current data when assuming different signal templates.
For the fiducial model template provided by Z12 and used in our models, reionization begins at $z=11$ and concludes at $z=8$. 
The patchy template scales with cosmology in the same way as hkSZ, shown in equation \ref{eq:ksz_scaling}.
 
 \subsection{Cosmic infrared background}
 \label{sec:cib}

The CIB is made up of thermal emission from dust grains which absorb optical and UV light and emit infrared radiation. 
The total energy in this emission is almost equal in power to all of the starlight that has not been absorbed \citep{dwek98, fixsen98}. 
The CIB signal originates in DSFGs over a large range in redshift \citep{lagache05,marsden09,casey14}. 
Models of the CIB are naturally expressed in units of flux density (Jy) as opposed to CMB temperature units. 
In order to convert between the power at an arbitrary frequency $\nu_0$ and the power in a cross spectrum $\nu_1 \times \nu_2$, we multiply the ratio of the flux densities by a conversion factor given by:
\begin{equation}
\epsilon_{\nu_1,\nu_2} \equiv \frac{\frac{dB}{dT}|_{\nu_0} \frac{dB}{dT}|_{\nu_0}}{\frac{dB}{dT}|_{\nu_1} \frac{dB}{dT}|_{\nu_2}},
\end{equation}
where $B$ is the CMB black-body specific intensity evaluated at \Tcmb, and $\nu_1$ and $\nu_2$ are the effective frequencies of the SPT bands, which are discussed in section Section \ref{sec:efffreq}. 

In this analysis, we separate the CIB term into Poisson and clustered components, allow each of these contributions to have independent frequency scaling, and allow a correlation between CIB
emission and the tSZ signal.
These portions of the CIB model are described in the next few sections.
\subsubsection{Poisson term} 
\label{sec:cibpois}

DSFGs appear as point sources in the SPT maps.
Statistical fluctuations in their number over the sky results in a Poisson distribution, which is a constant power in $C_\ell$, or $D_\ell \propto \ell^2$. 
The residual CIB power in the SPT power spectrum changes little with the flux threshold for point source masking, as the majority of power in the CIB is expected to come from sources below $\sim 1$ mJy. 

The Poisson power spectrum of the DSFGs can be written as
\begin{equation}
D^{p}_{\ell,\nu_1,\nu_2} = D^{p}_{3000} \epsilon_{\nu_1,\nu_2} \frac{\eta_{\nu_1} \eta_{\nu_2} } {\eta_{\nu_0} \eta_{\nu_0}}  \left( \frac{\ell}{3000}\right)^2,
\end{equation}
where $\eta_{\nu} $ describes how the CIB brightness scales with observing frequency, which is discussed in Section \ref{sec:cibfreqdep}. 
$D^{p}_{3000}$ is the amplitude of the Poisson DSFG power spectrum at $\ell=3000$ and frequency $\nu_0 = 154.1$\,GHz (Section \ref{sec:efffreq}). 
In this paper, we neglect variation in the frequency dependence between sources. 
\citet{hall10}  argue that the spectral index of the individual DSFGs varies by less than $\sigma_\alpha=0.7$, which would not substantially affect the Poisson power measured across SPT frequencies.

\subsubsection{Clustered term}
\label{sec:cibclus}

Galaxies are preferentially found within matter over-densities, which means they cluster on the sky. 
Previous studies of galaxies at sub-millimeter and millimeter-wavelength, including \citet{viero09}, \citet{dunkley11}, \citet{planck11-6.6_arxiv}, \citet{addison12a}, S11, and R12, have found the clustered term to be well-approximated by a  power law  of the form $D_\ell^c \propto \ell^{0.8}$. 
This power law index also matches  the correlation function observed for Lyman-break galaxies at $z \sim 3$ \citep{giavalisco98,scott99}. 
We use this power-law template unless specifically noted.

We consider two alternate forms for the clustered term. 
The first takes 1 and 2-halo clustered templates from the best-fit halo model in \citet{viero13a}. 
The frequency scaling is assumed to be the same between each template, however, the relative amplitude of each template is a free parameter. 
We also examine allowing the power-law exponent to vary.
As the effect of freeing the power-law exponent is similar to the effect of having free parameters for the amplitudes 1- and 2-halo terms, we only report CIB constraints for the latter case.

\subsubsection{Frequency dependence}
\label{sec:cibfreqdep}
 
 As the baseline model for the CIB frequency dependence, we use a modified black-body (BB) model. The frequency dependence in a modified BB model is:
\begin{equation}
\eta_{\nu}  = \nu^\beta B_\nu(T),
\label{eq:cibfreqdep}
\end{equation}
where $B_\nu(T)$ is the black-body spectrum for temperature $T$, and $\beta$ is an effective dust emissivity index. 
In the MCMC chains, we allow $\beta$ to be a free parameter with a uniform prior $\beta \in [0.5,3.5]$. 
Recent observations of galactic dust over a large portion of the sky have found $1 < \beta < 4$ in various regions \citep{bracco11}, and laboratory measurements of dust grain analogs show that $\beta$ is dependent on temperature and frequency and can span a large range. \citep{boudet05, agladze96, mennella98}
We generally fix $T = 20\,\mathrm{K}$, which
differs from the choice of \citet{addison12a} who fix the temperature to 9.7\,K.  
Since the SPT bandpasses are in the Rayleigh-Jeans region of the spectrum for any temperatures above $\sim 15$ K, there is a nearly perfect degeneracy between dust temperatures above this value. 
We allow $T$ to float between 5 and 50\,K in Section \ref{sec:cibinterp} and find virtually no impact on the SZ constraints.

We allow the  Poisson and clustered components of the CIB to have different frequency scalings ($\beta$). 
This choice is motivated by the different expected redshift distributions of the two components.
For fixed dust temperature, emission from low-redshift DSFGs will be closer to the Rayleigh-Jeans region 
of the spectrum and have a steeper spectral index than high-redshift emission coming from closer to the peak of the graybody.  
Large scale structure clustering is stronger at low redshift, so we expect the clustered component to have lower mean redshift, and hence a steeper frequency scaling, than the Poisson component \citep[e.g.,][]{addison12b}.
 
\subsubsection{tSZ-CIB correlations}
\label{sec:tszcibcorrel}

Both the tSZ and CIB are biased tracers of the underlying dark matter distribution, thus we expect
correlation between the two signals \citep{addison12c}. 
The level and angular dependence of the correlation are largely unconstrained by current observational data. 
The current \sptsz{} bandpowers show a clear preference for tSZ-CIB correlation (see Table~\ref{tab:deltachisq}), 
and a parameter describing this correlation is included in the baseline model.

We adopt a simple model that assumes the correlation is independent of $\ell$ and observing frequency. 
Below the tSZ null frequency (and, hence, in all SPT bands), 
the power from the tSZ-CIB correlation in this model takes the form:
\begin{equation}
D^{\mathrm{tSZ-CIB}}_{\ell,\nu_1,\nu_2}  = \nonumber
\end{equation}
\begin{equation}
\label{eqn:tszcib}
-\xi \left( \sqrt(D^{\mathrm{tSZ}}_{\ell,\nu_1,\nu_1}  D^{\mathrm{CIB}}_{\ell,\nu_2,\nu_2} ) +   \sqrt(D^{\mathrm{tSZ}}_{\ell,\nu_2,\nu_2}  D^{\mathrm{CIB}}_{\ell,\nu_1,\nu_1})   \right),
\end{equation}
where $D^{\mathrm{tSZ}}_{\ell,\nu_1,\nu_1}$ is the tSZ power spectrum, and $D^{\mathrm{CIB}}_{\ell,\nu_1,\nu_1}$ is the sum of the Poisson and clustered CIB components. 
Note that this is the opposite of the sign convention adopted in R12.

We consider one variation upon this simple model. 
When combining the tSZ model of \citet{shaw10} with the halo-model-based CIB models of \citet{shang12}, Z12 see an angular dependence in the correlation factor for some of the CIB models. 
Around $\ell = 3000$, the different correlation functions in Figure~3 of Z12 primarily vary in their slope after letting the amplitude float freely.
We thus consider an extension with one additional free parameter, a slope in $\ell$, such that:
\begin{equation}
\label{eqn:tszcib-slope}
\xi(\ell) = \xi + \xi^{slope} \frac{\ell}{3000}.
\end{equation}
The results for the tSZ-CIB correlation are discussed in Section \ref{sec:tszcibinterp}.

\subsection{Radio galaxies}
\label{sec:radio}

Most of the brightest point sources in the SPT maps are identified as radio sources in long-wavelength catalogs (e.g., SUMSS, \citealt{mauch03}) and have spectral indices that are consistent with synchrotron emission \citep{mocanu13}. 
Therefore, we include a Poisson term for radio sources below the detection threshold at 150 GHz of $\sim$6.4\,mJy:
\begin{equation}
D^{r}_{\ell,\nu_1,\nu_2} = D_{3000}^r \epsilon_{\nu_1,\nu_2} \left(\frac{\nu_1 \nu_2}{\nu_0^2}\right)^{\alpha_{r} }\left(\frac{\ell}{3000}\right)^2.
\end{equation}
 $D_{3000}^r$ is the amplitude of the Poisson radio source power spectrum at $\ell=3000$ and frequency $\nu_0$. 
 We denote the effective spectral index of the radio source population as $\alpha_r$. 
 $D_{3000}^r$ will depend linearly on the flux above which point sources are masked as a consequence the fact that $S^2 dN/dS$ for synchrotron sources is nearly independent of $S$ (see e.g., \citealt{dezotti05}).

We do not expect to strongly detect the Poisson signal at 150\,GHz from sources below the SPT 
detection threshold; instead, we apply a prior to this term based on models.
As in R12,  a Gaussian prior is set on  the 150\,GHz amplitude of $1.28 \pm 0.19 \,\uksq{}$ based on the \citet{dezotti05} source count model. 
We note that this model provides a reasonable fit to the measured SPT 150\,GHz synchrotron-dominated source
counts in the flux range near the detection threshold \citep{mocanu13}.
R12 also fixed the spectral index to -0.53---effectively setting a prior on the radio Poisson signal at 
95 and 220\,GHz with the same fractional width as the 150\,GHz prior---because freeing the spectral index did not affect the SZ constraints. 
With the improved data in this work, we have a strong detection of radio power at 95\,GHz, and so we 
allow the spectral index to vary with a uniform prior on $\alpha_r \in [-2,0]$, which frees the 95 GHz power from a prior on the source amplitude.
Fixing the spectral index to $\alpha_r = -0.53$ as in R12 reduces the uncertainties on \dtsz{} and \dksz{} by $\sim$\,10\% and shifts the median values by $\sim$\,0.5\,$\sigma$. 
We tested removing the radio prior, and find that slight changes to the prior on the 150\,GHz amplitude has a negligible effect on the SZ parameters. 
When the prior is removed however, we have no detection of the 150 GHz radio amplitude.
We therefore conclude that the 150 GHz amplitude prior is well-motivated, and include it in all chains.

We expect the clustering of radio galaxies to be negligible at SPT frequencies. 
First, \citet{hall10} show that clustering on the relevant scales will be less than 5\% of the mean, which is already quite low compared to the CIB. 
Additionally, clustered signals at 30 GHz \citep{sharp10} extrapolated to 150 GHz would contribute less than 0.1\,\uksq{}, well below the uncertainty on the clustered CIB signal at that frequency.
We therefore ignore radio source clustering in this work.

We do not expect significant correlations between radio galaxies and the tSZ signal.
\citet{lin09} and \citet{sehgal10} show that the number of radio sources with significant mm wavelength flux that are correlated with galaxy clusters is expected to be low.
\citet{shaw10} produced a cross spectra of the simulated tSZ and radio source maps produced in \citet{sehgal10} after masking sources above 6.4 mJy.
\citet{shaw10} found that the correlation coefficient of radio galaxies and tSZ was 0.023 in these masked simulated maps, which agrees with other lines of reasoning for a low tSZ-radio correlation. 
If we turn this into a power contribution in the most sensitive 150\,GHz band, the tSZ-radio correlation term would be approximately 0.1\,\uksq, or 1/15 that of the tSZ-CIB term. 
We therefore ignore tSZ-radio correlations in this work.

\subsection{Galactic cirrus}
\label{sec:cirrus}

Galactic cirrus is the final and smallest foreground term included in the model. 
We marginalize over the cirrus using a fixed angular template with a frequency dependence described by a modified BB. 
To determine the angular dependence and priors on the cirrus power at each frequency, we cross-correlate the SPT maps with the model 8 galactic dust predictions of \cite{finkbeiner99}. 
Galactic cirrus with an angular dependence of
\begin{equation}
D^{\mathrm{cir}}_{\ell,\nu_1,\nu_2} = D_{3000}^{{\mathrm{cir}}, \nu_1,\nu_2} \left(\frac{\ell}{3000}\right)^{-1.2}
\end{equation} 
is seen in all frequency bands with signal to noise of $\sim 3\,\sigma$.
The measured powers are 0.16, 0.21, and $2.19\,\uksq{}$ at 95, 150, and 220\,GHz. 
The three cirrus parameters are the three auto-spectra powers, and the cross-frequency powers are the geometric means of these parameters.
We set Gaussian priors based on the measured power and uncertainty at each frequency, thus there are three priors and three parameters in the cirrus model. 
Fixing the cirrus power instead does not change the constraints.

\subsection{SPT effective frequencies}
\label{sec:efffreq}

We refer to the SPT frequencies as 95, 150, and 220\,GHz. 
The detailed bandpasses were measured using a Fourier transform spectrometer (FTS). 
The bandpasses changed slightly between 2008 and 2009 due to a focal plane refurbishment, and remained essentially unchanged thereafter. 
With the data from the FTS measurements, we can calculate an effective band center for sources with any spectrum. 
As we calibrate the data in CMB temperature units, the effective frequency does not matter for sources with a CMB-like spectrum. 
The effective band center can vary with the source spectrum, however, and is calculated for all other model components. 

We calculate the effective frequency for each year's spectral bandpass and average them according to the yearly weights (see Section \ref{sec:bandpowerest}). 
For an $\alpha = -0.5$ (radio-like) source spectrum, we find band centers of 95.3, 150.5, and $214.0\,$GHz. 
For an $\alpha = 3.5$ (dust-like) source spectrum, we find band centers of 97.9, 154.1, and $219.6\,$GHz. 
If we reduce $\alpha$ by 0.5, the band centers drop by $\sim 0.2$ GHz for both radio and dust-like specta; we neglect this small shift in the model fitting. 
We use these band centers to calculate the frequency scaling between the SPT bands for the radio and CIB terms. 
For a tSZ spectrum, we find band centers of 97.6, 153.1, and $218.1\,$GHz. 
The ratio of tSZ power in the 95\,GHz band to that in the $150\,$GHz band is 2.80. 
The 220 GHz band center is at the tSZ null, thus we expect no tSZ signal in the 220 GHz band.
Unless specifically noted, we will quote the tSZ power constraints at 143\,GHz for consistency with \planck.

\section{Results}
\label{sec:results}

\subsection{Baseline model}
\label{sec:baseres}

We begin by presenting results for the baseline model discussed in Section \ref{sec:model}. 
This model includes the six $\Lambda$CDM parameters plus eight parameters to describe foregrounds. 
Foreground parameters include the amplitudes of the tSZ power, kSZ power, CIB Poisson power, and CIB clustered power; two parameters to describe the frequency dependence of the CIB terms; the tSZ-CIB correlation; and the spectral index of radio galaxies. 
The amplitudes of the Poisson radio galaxy power at 150\,GHz and galactic cirrus are technically free
parameters but are constrained by strong priors.

\begin{figure*}[tbh]\centering
\includegraphics[trim=1.9cm 2.5cm 9.5cm 8cm,clip,width=0.9\textwidth]{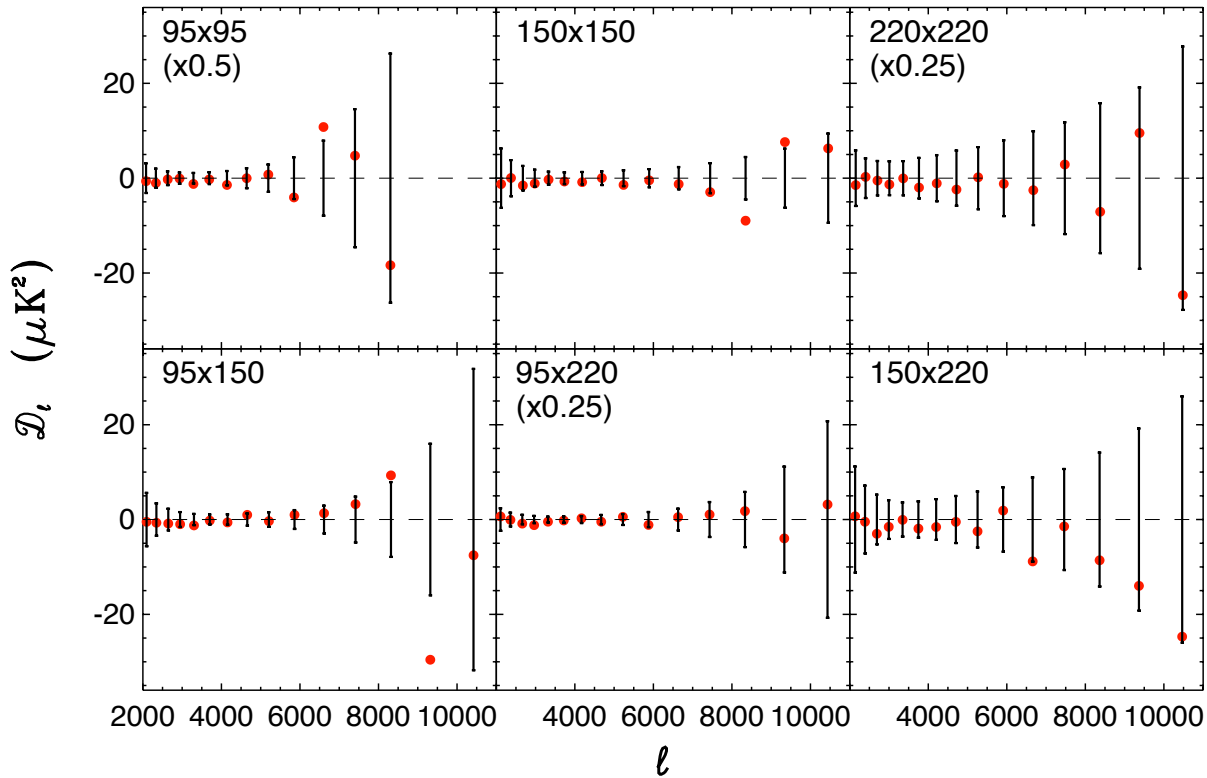}
\caption[]{ 
Best-fit residuals on the six auto- and cross-spectra measured with the 95, 150, and 220\,GHz SPT data. 
The residuals have been corrected by the best-fit calibration, but no beam uncertainties have been included. 
For plotting purposes, the residuals and error bars of three spectra have been renormalized by factors of 0.25-0.5 (marked on plot). 
  }
  \label{fig:resid}
\end{figure*}

We fit the 88 SPT bandpowers to the model described above. 
There are 80 degrees of freedom (dof), since the $\Lambda$CDM parameters are essentially fixed by the \planck{}+BAO+\ho{} data, leaving the eight foreground model parameters. 
Fixing the $\Lambda$CDM parameters to their best-fit values has little effect on derived constraints. 
This baseline model fits the SPT data with \chisq{} = 89.7 and a PTE of 21.5\%, and provides the simplest interpretation of the data. 
The residuals are shown in Figure~\ref{fig:resid}, with the best-fit calibration (but not beam errors) applied. 

Table~\ref{tab:deltachisq} shows the improvement in the quality of the fits with the sequential
introduction of free parameters to the original $\Lambda$CDM primary CMB model.
 Adding additional free parameters beyond a term for clustered DSFGs does not significantly improve the quality of the fits.

\begin{table}[ht!]
\begin{center}
\caption{\label{tab:deltachisq} Delta $\chi^2$ for model components}
\small
\begin{tabular}{cc|cc}
\hline\hline
\rule[-2mm]{0mm}{6mm}
term & dof & \delchisq\  \\
\hline
CMB + Cirrus &- & (reference)\\
DSFG Poisson&2 & -2383.8\\
Radio Poisson& 1 & -555.5\\
tSZ & 1 &  -263.2\\
DSFG Clustering&2&  -199.1\\ 
 tSZ-CIB Correlation & 1 &  -1.5\\
  kSZ & 1 &  -0.9\\
\hline
\hline
  1/2 halo clustered DSFG model & 1 &  -1.8 \\
    Sloped tSZ-CIB corr. & 1 &  -0.9\\
      $T\in[5,50\,K]$ & 2 & -0.2 \\
\hline
\end{tabular}
\tablecomments{ Improvement to the best-fit $\chi^2$ as additional terms are added to the model. 
Terms above the double line are included in the baseline model, with each row showing the improvement in likelihood relative to the row above it. 
For rows below the double line, the $\Delta\chi^2$ is shown relative to the baseline model rather than the row above it. 
The row labeled ``1/2 halo clustered DSFG model " replaces the power-law template by two templates for the 1- and 2-halo terms from the halo model in \citet{viero13a}. 
The linear combination of the two templates relaxes the shape prior on the clustered DSFG term. 
This extension has the strongest support in the data. 
The row labeled ``Sloped tSZ-CIB corr." relaxes the assumption that the tSZ-CIB correlation is independent of angular scale, allowing it to vary linearly with $\ell$. 
The row labeled ``$T\in[5,50\,K]$" allows the temperature of the modified BB for the Poisson and clustered CIB terms to vary between 5 and 50\,K. 
} \normalsize
\end{center}
\end{table}

\subsubsection{CIB constraints}
\label{sec:cibbaseres}

Both the Poisson and clustered CIB components are detected at high significance. 
At 150 GHz, we find that the Poisson component has  power $D^{p}_{3000}=9.16 \pm 0.36\, \uksq{}$ and the clustered component has power $D^{c}_{3000}=3.46 \pm 0.54 \,\uksq{}$ at $\ell = 3000$. 
 At 220\,GHz, we measure  $D^{p, 220x220}_{3000}=66.1 \pm 3.1\, \uksq{}$ and $D^{c, 220x220}_{3000}=50.6 \pm 4.4\, \uksq{}$. 
The parameter $\beta$ in equation \ref{eq:cibfreqdep} is constrained to be $1.505 \pm 0.077$ for the Poisson term and $2.51 \pm 0.20$ for the clustered CIB term. 
This translates to effective spectral indices between 150 and 220 GHz of $3.267\pm0.077$ and $4.27\pm0.20$ for the Poisson and clustered terms respectively. 

The constraints from the baseline model are roughly in line with both theoretical expectations and previous work (\citet{hall10}, S11, \citet{dunkley11}, R12), though we do see less clustered power than previous observations. 
With tSZ-CIB correlations, R12 found the Poisson power at 150\,GHz to be $D^{p}_{3000}=8.04 \pm 0.48 \, \uksq{}$, and constrained the clustered power at 150\,GHz to be $D^{c}_{3000}=6.71 \pm 0.74 \, \uksq{}$.
Most of the difference in clustered power is due to the different foreground modeling in R12. 
Fitting the R12 data to the baseline model in this work leads to $D^{p}_{3000}= 8.59 \pm0.63$ and $D^{c}_{3000} = 4.00\pm1.1$ at 150 GHz, which represent shifts of less than 1$\sigma$ between the two data sets. 
The model change that affected the clustered power constraint the most is allowing both a tSZ-CIB correlation and independent spectral indices for the Poisson and clustered CIB simultaneously.
The two model parameters were considered separately as extensions to the baseline model in R12, but were not considered together.
Additionally, the more precise 90 GHz data presented here constrain the radio spectral index to be slightly lower than the fixed value adopted in R12 (see \S\ref{sec:rgbaseres}). 
The change in radio spectral index also contributes a small amount to the change in clustered CIB power.

The implications of the CIB constraints are discussed in Section \ref{sec:cibinterp}.
 
 \subsubsection{Radio galaxy constraints}
\label{sec:rgbaseres}

Radio power is detected at high significance in the 95\,GHz data: 
$D^{r-95 \times 95}_{3000}=7.81 \pm 0.75 \, \uksq{}$. The data show a mild preference
for lower radio power at 150\,GHz than assumed by the prior: 
$D^{r-150\times 150}_{3000}=1.06 \pm 0.17 \, \uksq{}$, or 0.9\,$\sigma$ lower than the prior 
value of $1.28 \pm 0.19 \, \uksq{}$. Taken at face value, this implies a 
preferred radio spectral index of $\alpha_{r} = -0.90\pm 0.20$,
which is approximately 1.5\,$\sigma$ lower than the median spectral index of -0.60 for synchrotron-classified sources 
in the SPT survey, as reported in \citet{mocanu13}. 
A lower spectral index is consistent with the picture that the spectral index flattens for the brightest 150\,GHz radio sources, as argued by \citet{mocanu13}.  
It could also be caused in part by the selection of point sources masked in this power spectrum analysis: sources detected at 150\,GHz at a signal-to-noise $>$ 5 are masked. 
A source with a given 95\,GHz flux is more likely to remain unmasked if it has a steep spectral index. 
For the purposes of this work, however, the radio spectral index is a nuisance parameter, and 
a detailed study of the relative importance of these effects is beyond the scope of this paper.
 
\subsubsection{SZ power} 
\label{sec:ksztszbaseres}

\begin{figure}[htb]
\begin{center}
\resizebox{0.45\textwidth}{!}{
\includegraphics[width=1.0\textwidth,clip=true, trim = 1.15cm 11.35cm 8.16cm 4.09cm]{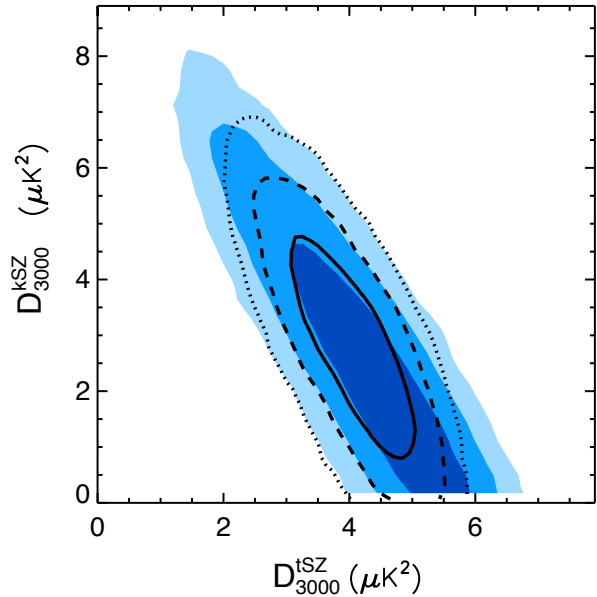}}
\end{center}
\caption[
2d likelihood of tSZ and kSZ]{2D likelihood surface for the tSZ and kSZ power at 143 GHz at $\ell=3000$ in the baseline model including tSZ-CIB correlations. 
1, 2, and 3 $\sigma$ constraints are shown in shades of blue for the current data, and by the solid, dashed, and dotted lines respectively when the bispectrum prior is included.
The observed degeneracy is due to the potential correlation between the tSZ and CIB.
}
\label{fig:1paneltszksz}
\end{figure}

As shown in Figure~\ref{fig:1paneltszksz}, we strongly detect tSZ power and place an upper limit on the kSZ power. 
We measure $\dtsz=4.38^{+0.83}_{-1.04}\,\uksq{}$ and set a 95\% CL upper limit on $\dksz < 5.4\,\uksq{}$. 
The current data best constrain a linear combination of the two powers of $\dtsz + 0.55 \dksz = 5.66 \pm 0.40 \,\uksq{}$. 
 The degeneracy axis of the tSZ and kSZ powers is slightly tilted compared to R12. 
 We note that this shift in the degeneracy axis is largely due to reporting tSZ power at 143\,GHz rather than $153$\,GHz as in previous work.
SZ power is detected at approximately 14\,$\sigma$. 
These results are nearly independent of the detailed tSZ and kSZ template shapes. 
Table \ref{tab:szconstraint} shows the SZ power constraints for different templates used in the modeling.
The different tSZ templates are discussed in Section \ref{sec:tszinterp}, while the different kSZ templates are discussed in Section \ref{sec:kszinterp}.

These SZ measurements are consistent with earlier observations of the SZ power. 
R12 reported tSZ power at 152.9\,GHz;  numbers at 143\,GHz will be a factor of 1.288 higher. 
Scaling to 143\,GHz, the result reported by R12  for a model with a free tSZ-CIB correlation is $\dtsz=4.20 \pm 1.37$.
This is a shift of $0.1\,\sigma$. 
R12 also found a consistent (if weaker) 95\% CL upper limit on the kSZ power of $\dksz < 6.7 \,\uksq{}$. 
SZ power has also been constrained with ACT data, most recently by \citet{dunkley13} at 150\,GHz. 
Translating to 143\,GHz, the ACT constraints of $\dtsz=3.9 \pm 1.7 \,\uksq{}$ and $\dksz < 8.4 \,\uksq{}$ are consistent with those presented here. 
The results presented in this work are consistent with previous analyses of the small-scale SZ power spectra with significantly smaller uncertainties.

\begin{table*}[ht!]
\begin{center}
\caption{\label{tab:szconstraint} SZ constraints}
\small
\begin{tabular}{cc|c|c|c}
\hline\hline
\rule[-2mm]{0mm}{6mm}
tSZ Prior& kSZ Template  & $D^{\rm tSZ}_{3000} ~(\mu {\rm K}^2)$ & $D^{\rm kSZ}_{3000} ~(\mu {\rm K}^2)$ & $\xi$ \\
\hline
- &CSF+patchy  & $ 4.38 \pm  0.94$ & $<  5.4$ & $0.100 \pm  0.061$\\
- &CSF  & $ 4.63 \pm  0.86$ & $<  4.5$ & $0.085 \pm  0.054$\\
- &Patchy  & $ 4.09 \pm  1.03$ & $<  6.6$ & $0.121 \pm  0.072$\\
Bispec. &CSF+patchy  & $ 4.08 \pm  0.63$ & $<  4.9$ & $0.113 \pm  0.054$\\
Bispec. &CSF  & $ 4.18 \pm  0.61$ & $<  4.4$ & $0.103 \pm  0.053$\\
Bispec. &Patchy  & $ 3.95 \pm  0.63$ & $<  5.5$ & $0.127 \pm  0.057$\\
\hline
- &(Fixed CSF) + Patchy  & $ 3.58 \pm  0.76$ & $<  4.4$ & $0.152 \pm  0.065$\\
Bispec. &(Fixed CSF) + Patchy  & $ 3.74 \pm  0.52$ & $<  3.3$ & $0.141 \pm  0.051$\\
\hline
\end{tabular}
\tablecomments{ 
Measured tSZ power and tSZ-CIB correlation, and 95\% confidence upper limits on the kSZ power at $\ell=3000$.  
The Shaw tSZ model template is used in all rows. 
The first row has constraints for the fiducial model. 
The second and third rows demonstrate how the constraints depend on the kSZ template, specifically when all kSZ power is assumed to be in the form of the CSF homogeneous kSZ template or patchy kSZ template. 
The real kSZ spectrum should be a mixture of the two kSZ templates as in row 1. 
The fourth through sixth rows follow the same pattern, but include the bispectrum information.
Below the solid line in rows seven and eight, we show the SZ constraints when we fix the homogeneous kSZ to the CSF model and allow a free parameter for additional patchy kSZ.
The CSF kSZ model predicts $1.57\,\mu{\rm K}^2$ for the WMAP7 cosmology detailed in Section \ref{sec:hksz} and is scaled with cosmology according to the prescription in that section. 
The kSZ power shown in these rows is expected to be entirely from the epoch of reionization. 
}
\normalsize
\end{center}
\end{table*}

\subsubsection{tSZ-CIB correlation} 
\label{sec:tszcibbaseres}

\begin{figure*}[htb]\centering
\begin{center}
\resizebox{0.8\textwidth}{!}{
\includegraphics[width=1.0\textwidth,clip=true, trim =  2.06cm  11.97cm  9.32cm 10.22cm]{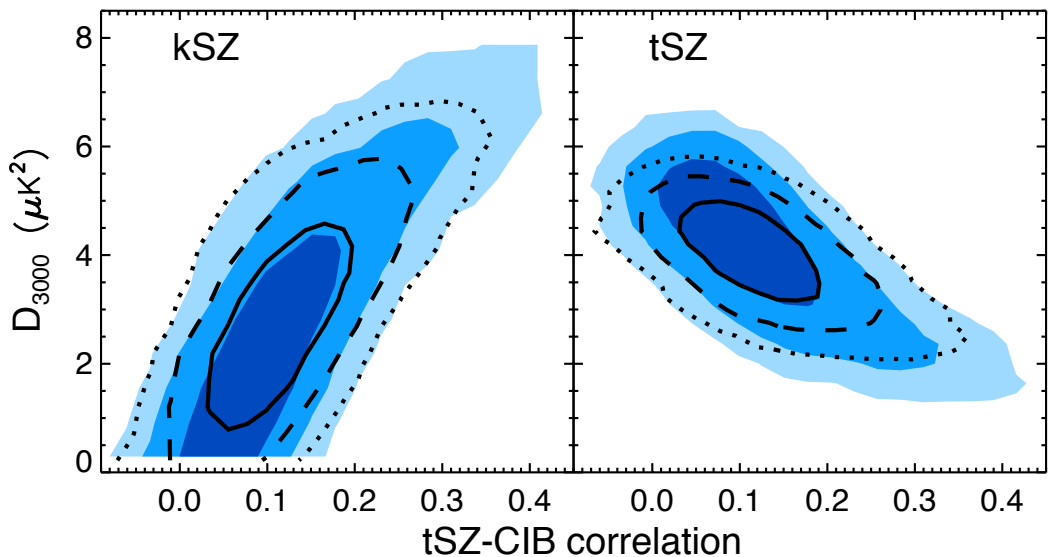}}
\end{center}
\caption[
tSZ-CIB correlation 2d likelihood with kSZ and tSZ.]{
The  2d likelihood of the tSZ-CIB correlation and kSZ (\textbf{\textit{Left panel}}) or tSZ power (\textbf{\textit{Right panel}}). 
The filled contours show the 1, 2, and $3\,\sigma$ constraints. 
The line contours show the same with the addition of the bispectrum prior on tSZ power. 
The data strongly prefer a positive correlation, consistent with DSFGs being over-dense in galaxy clusters. 
}
\label{fig:2paneltszcib}
\end{figure*}

We  parameterize the tSZ-CIB correlation with a single parameter $\xi$. 
An overdensity of dusty galaxies in galaxy clusters would result in a positive value of $\xi$. 
The tSZ-CIB correlation is partially degenerate with the tSZ and kSZ power, as illustrated in  
Figure \ref{fig:2paneltszcib}. 
Increasing the correlation  decreases tSZ power and increases kSZ power. 
We measure the tSZ-CIB correlation to be $\xi = 0.100^{+0.069}_{-0.055}$, consistent with the $\xi = 0.18 \pm 0.12$ measured by R12. 
The data prefer positive tSZ-CIB correlation, and rule out $\xi <0$ at the 98.1\% CL.

\subsection{Robustness to cosmology} 
\label{sec:beyondlcdm}

A natural question is the extent to which the reported SZ and CIB constraints depend on the assumption of a standard \lcdm{} cosmology in calculating the primary CMB temperature power spectrum. 
We examine two extensions to \lcdm: the number of relativistic species (\neff) and massive neutrinos (\summnu). 
These extensions have been selected as a representative sample of extensions that significantly affect the shape of the primary CMB damping tail, and thus might influence the derived SZ or CIB constraints. 
We also test fixing the tensor-to-scalar ratio $r$ to 0.2 instead of 0, motivated by the recent BICEP2 results \citep{bicep2a}. 
We find minimal impact on  the SZ and foreground parameters. 
The largest shift in any of the parameters with one of these extensions is $0.3\,\sigma$ (0.1\,\uksq{}) to the DSFG Poisson power level. 
We also test whether {\em reducing} freedom in the primary CMB temperature power spectrum model impacts the SZ constraints.  
We find fixing the primary CMB temperature power spectrum to the best-fit model results in negligible changes. 

Additionally, we have checked to ensure that the calibration of the bandpowers using \wmap{} versus \planck{} does not affect the results. The SPT calibration shift is 2-3\% in power between \wmap{} and \planck{}, depending on the frequency. None of the foreground or SZ model parameters that this paper focuses on, however, are measured with percent level accuracy. The observed shifts for the parameters across the three bands are well below $1 \sigma$.

We conclude that, given the high precision with which the primary CMB power spectrum has been measured and its radically different $\ell$-dependence, the SZ and foreground constraints are insensitive to the remaining uncertainty in the primary CMB temperature anisotropy.

\subsection{Constraints from the tSZ bispectrum} 
\label{sec:withbispec}

More information about the tSZ effect can be gleaned from higher-order moments of the map, such as the bispectrum. 
The bispectrum, and other higher-order statistics, are less affected by the kSZ-tSZ degeneracy because the kSZ effect is nearly Gaussian and thus contributes very little to the bispectrum. 
Adding bispectrum measurements can partially break that degeneracy, subject to modeling uncertainties in going from the bispectrum to power spectrum. 
Motivated by this potential, we investigate adding an external  tSZ power constraint derived from the 800 deg$^2$ SPT bispectrum measurement in C14. 
At 153.8\,GHz, C14 report $D_{3000}^\mathrm{tSZ-153.8\,GHz} = 2.96 \pm 0.64 \,\mu$K$^2$; this translates to $\dtsz = 3.81 \pm 0.82 \,\mu$K$^2$ at 143\,GHz.

As would be expected, the  bispectrum prior primarily impacts the SZ constraints, with a small impact on the amplitude of the tSZ-CIB correlation. 
There is no impact on the CIB constraints, with the largest shift being only 0.2\,$\sigma$. 
There is a small downward shift (0.3\,$\sigma$) in the median tSZ power from $\dtsz=4.38^{+0.83}_{-1.04} \ \,\uksq{}$ to $4.08^{+0.58}_{-0.67} \,\uksq{}$, and a 30\% reduction in the uncertainties. 
This in turn pushes the ML kSZ power higher and, when relaxing the positivity prior by running chains with negative values of kSZ power allowed, we find $\dksz{}= 2.87 \pm 1.25\,\uksq{}$. 
This translates to a 95\% CL upper limit of 4.9\,\uksq, and disfavors negative kSZ power at 98.1\% CL. 
When including the bispectrum measurement, the magnitude of the tSZ-CIB correlation is increased, going from $\xi = 0.100^{+0.069}_{-0.053}$ to $\xi = 0.113^{+0.057}_{-0.054}$.
This rejects anti-correlation ($\xi<0$) at 99.0\% CL. 

\subsection{Results for cluster masked bandpowers} 
\label{sec:clustermask}

We finally consider the effect of masking all galaxy clusters detected at $>5$\,$\sigma$ in the \sptsz{} survey
(Bleem et al., in preparation).
Measuring the tSZ power spectrum with clusters masked is an observational probe of how much tSZ power is coming from clusters above the detection threshold of the cluster survey.
The masking of the galaxy clusters mirrors the point source treatment with a disk radius of 5\arcmin{} and then a Gaussian taper with FWHM =5\arcmin{}. 
The disk radius is chosen to be at least a factor of two larger than the core radii of  galaxy clusters expected to contribute strongly to the tSZ power at $\ell=3000$. 
The bandpowers for this mask are available online.$^{\ref{footnote:spt}}$
We fit these bandpowers to the baseline model. 
As expected, masking galaxy clusters reduces the  measured tSZ power significantly. 
The masked value of \dtsz\ shifts by 2.05\,\uksq{} (1.5\,$\sigma$) from $4.38^{+0.83}_{-1.04}$\,\uksq{} to $2.33^{+0.80}_{-1.00}$\,\uksq{}. 
Unsurprisingly, the next most significant shift is to the tSZ-CIB correlation, the magnitude of which increases  from $\xi = 0.100^{+0.069}_{-0.053}$ to $\xi = 0.155^{+0.137}_{-0.080}$. 
The increased uncertainty is a natural consequence of decreasing the tSZ power. 
The overall shift in the tSZ-CIB correlation could be caused by increased emission from DSFGs relative to the tSZ signal in less massive clusters. 
A detailed analysis of this effect, however, is beyond the scope of this work. 
The constraints on the kSZ power and other foreground terms are essentially unchanged; the 95\% CL upper limit on the kSZ power goes from 5.4 to 5.5\,\uksq. 

Using symmetrized errors for the cluster-masked and default values of tSZ power, the ratio of power
is $0.53 \pm 0.24$. We can compare this value to predictions in the literature for the fraction of tSZ power from clusters above or below a given mass. 
Using mass estimates from the \sptsz{} cluster catalog (Bleem et al., in preparation), we find that a detection significance of 5$\sigma$ closely corresponds to a mass of $M_{500} = 2 \times 10^{14} M_\odot / h$, where $M_{500}$ is the mass enclosed within the radius $R_{500}$, within which the cluster density is equal to 500 times the critical density of the universe.
Figure 3 of \citet{bhattacharya12} shows that, assuming the \citet{tinker08} mass function and the \citet{arnaud10} pressure profile, between $50\%$ and $60\%$ of the tSZ power spectrum at $\ell=3000$ comes from clusters less massive than $M_{500} = 2 \times 10^{14} M_\odot / h$. 
The measured value falls in the middle of this range. 
Qualitatively similar behavior is shown in Figure 6 of \citet{komatsu02} and Figure 1 of \citet{holder02b}, which use different halo mass functions and cluster physics models.
A detailed comparison of not just the amplitude but also the shape of the tSZ power spectrum as a function of cluster mass cut could potentially provide interesting constraints on the pressure profiles of galaxy clusters \citep{battaglia12}, however, such an analysis is outside the scope of this work.

\section{tSZ interpretation} 
\label{sec:tszinterp}

\begin{figure}[thb]\centering
\includegraphics[width=0.45\textwidth,clip,trim =  2.47cm  11.84cm  10.0cm  6.47cm
]{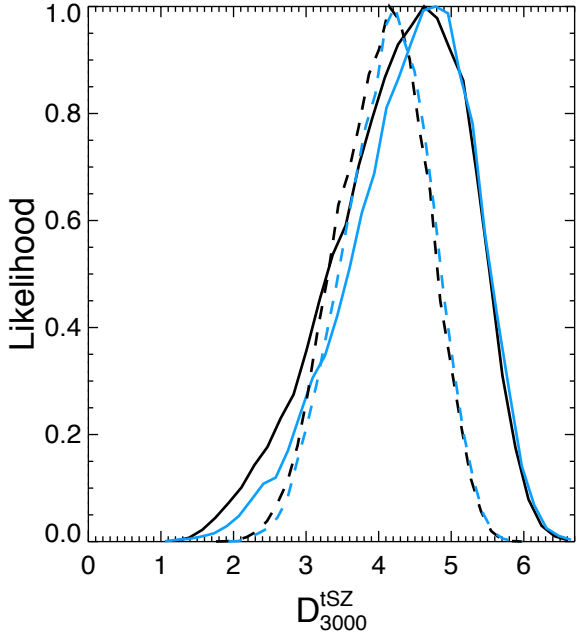}
  \caption[]{
  1D likelihood curves for the amplitude of the tSZ power spectra at 143\,GHz and $\ell=3000$. 
  The \textbf{solid} and \textbf{dashed} lines are without and with the tSZ bispectrum information. 
  The \textbf{black} curves are for the fiducial model; the nearly identical \textbf{blue} curves show the likelihood when a freely varying slope in the tSZ-CIB correlation is introduced. 
  }
   \label{fig:1dtsz}
\end{figure}

In this section, we consider the implications of the observed tSZ power.
The tSZ power constraints are reported in Section \ref{sec:ksztszbaseres} and Table \ref{tab:szconstraint}, and the 1D likelihoods for the baseline model and one extension are shown in Figure \ref{fig:1dtsz}. 
We first combine  our measurement of the tSZ power spectrum with measurements of the primary CMB power spectrum, H$_0$, and BAO to constrain cosmological constants, in particular $\sigma_8$. 
We then discuss the limitations on cosmological constraints due to uncertainty in theoretical predictions for the tSZ power spectrum and the tension between these predictions and our measurements, given current constraints on cosmology.

\begin{table}[ht!]
\begin{center}
\caption{\label{tab:szchisq} Delta $\chi^2$ for SZ models}
\small
\begin{tabular}{c|c}
\hline\hline
\rule[-2mm]{0mm}{6mm}
tSZ Model & \delchisq{}  \\
\hline
{\it Free} Shaw & - \\
Bhattacharya & +4.3 \\
Shaw & +6.8 \\
Sehgal & +38.8 \\
\hline
\end{tabular}
\tablecomments{ 
Change to the best-fit $\chi^2$ when the tSZ power spectrum
  is fixed to the predictions of different models.
  Differences are reported relative to the baseline model with a free amplitude for the Shaw tSZ template.
  The model predictions are scaled to account for the specific
  cosmology at each chain step as described in Section \ref{sec:tsz}.
  The kSZ power is unconstrained in each case. 
     } \normalsize
\end{center}
\end{table}

\begin{figure}[thb]\centering
\includegraphics[width=0.45\textwidth,clip,trim =  2.47cm  11.84cm  10.0cm  6.47cm]{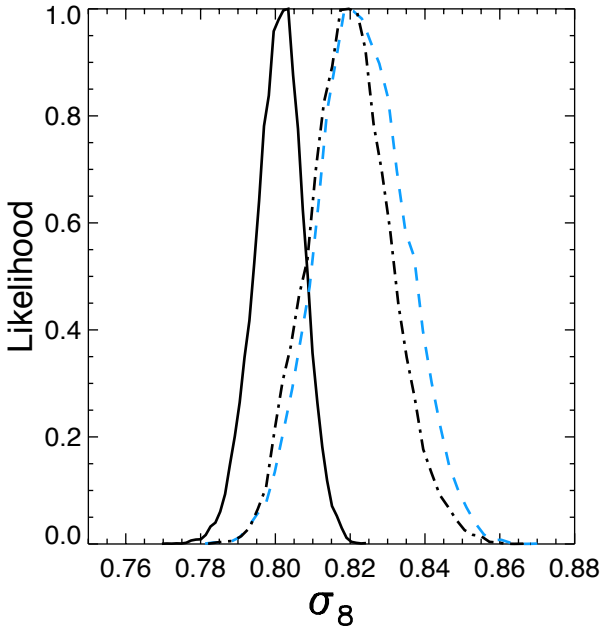}
  \caption[]{
   1D likelihood curves for $\sigma_8$. 
   The \textbf{dashed blue} line marks the constraints with no tSZ information, from primary CMB + BAO + H$_0$ alone. 
   The \textbf{dot-dashed black} line shows the constraint with the tSZ information based on the Bhattacharya tSZ model with a 50\% modelling uncertainty. 
   The \textbf{solid black} line is the same except with no modelling uncertainty.  
  }
   \label{fig:sigma8}
\end{figure}

The raw statistical power of the tSZ power spectrum in constraining cosmological parameters---particularly
$\sigma_8$---is impressive.
The very steep scaling shown in equation~\ref{eq:tsz_scaling} ($D^\mathrm{tSZ} \propto \sigma_8^{8.3}$)
implies that the 
measurement of $D^\mathrm{tSZ}$ presented in Section \ref{sec:results} would result in a statistical 
constraint on $\sigma_8$ of $\sigma(\sigma_8) = 0.0057$, or a fractional uncertainty of $< 1\%$.

To go from the observed tSZ power to a constraint on $\sigma_8$, however, requires a prediction for  the tSZ power as a function of cosmological parameters. 
This prediction is subject to considerable uncertainty, as evidenced by the wide spread in predictions for $D^\mathrm{tSZ}$ for a fixed cosmology by the three models discussed in Section \ref{sec:tsz}.
The difficulty in predicting the tSZ power spectrum lies partly in the fact that a significant fraction of the total tSZ power is contributed by groups and clusters of mass $M_{500} < 2\times 10^{14} \hmsun$ and redshift $z > 0.65$ \citep{trac11, battaglia12,shaw10}, a regime in which few observational constraints exist. 
This difficulty can be ameliorated somewhat by measurements of the tSZ bispectrum (three-point function), which probes an intermediate range of clusters in mass and redshift and can act as a bridge between the well-studied clusters and those that contribute primarily to the tSZ power spectrum \citep{bhattacharya12,hill13}.
Even for well-studied low-redshift, high-mass clusters, significant uncertainty exists in the amount of non-thermal pressure support in clusters, particularly in the outskirts \citep{shaw10,battaglia12a, trac11,parrish12}. 
It is primarily the level of non-thermal pressure support assumed in the three models in Section \ref{sec:tsz} (the Sehgal, Shaw, and Bhattacharya models) that causes the factor-of-two difference in predicted tSZ power at fixed cosmology.

Following previous SPT power spectrum publications, we assume a 50\% modeling uncertainty on the tSZ power spectrum at fixed cosmology---roughly encompassing the predictions of the three models we consider---and derive a constraint on $\sigma_8$. 
In principle, we would also need to estimate the non-Gaussian sample variance between the observed tSZ power and the true cosmological value, but in practice the sample variance is negligibly small compared to the model uncertainty. 
Based on the \citet{shaw09} simulations, L10 estimated the fractional uncertainty due to sample variance on the amplitude of the tSZ power spectrum to be 12\% for 100\,\sqdeg{}, implying an expected value of 2.4\% for the 2500\,\sqdeg{} of sky used here.
With these assumptions, we obtain a constraint of $\sigma_8 = 0.820 \pm 0.011$ for the Bhattacharya model and a similar constraint for the Shaw model, while the constraint assuming the Sehgal model as the central value of the prior is shifted slightly (less than $1 \sigma$ ) downward.
These constraints are only modestly shifted from the primary CMB+BAO+\ho{} constraint of $\sigma_8 = 0.823 \pm 0.012$, and the uncertainty is not significantly reduced.
 
These results demonstrate that, with a 50\% modeling uncertainty, the tSZ power spectrum does not add 
significant weight to the already tight constraints from primary CMB, BAO, and \ho{}. 
On the other hand, this implies that the existing cosmological constraints on $\sigma_8$ are placing strong constraints on the models of cluster physics. 
As shown in Table~\ref{tab:szchisq}, the central values of the relation between the tSZ power spectrum and cosmology for all three models we consider are in some tension with the combination of existing constraints on cosmology and the tSZ power we observe. 
If we were to interpret this as a constraint on a single cluster gas physics parameter such as the level of non-thermal pressure support (the primary difference between the three tSZ models), the Sehgal model value is ruled out at greater than $6 \sigma$, while the Shaw and Bhattacharya models values are disfavored at more than  $2 \sigma$, all in the direction of the data preferring larger values of non-thermal support.

Put another way, the values of $\sigma_8$ that would be derived from using tSZ measurements and adopting any of these models with no uncertainty would be significantly lower than the current best-fit \LCDM{} value from primary CMB+BAO+\ho{}. 
Figure \ref{fig:sigma8} shows the constraints on $\sigma_8$ both with and without modelling uncertainty, as compared to the $\sigma_8$ constraint from primary CMB+BAO+\ho{} alone.
This discrepancy has been noted in earlier tSZ measurements. 
For example, L10 found that the measurement of the tSZ power spectrum in that work, combined with assuming the Sehgal model with zero prior width, resulted in a constraint of $\sigma_8=0.746 \pm 0.017$. 
This tension is not confined to power spectrum measurements with the tSZ effect; cluster count measurements find a similar magnitude and direction of tension.
For example, \citet{planck13-20} used counts of tSZ-selected clusters, an X-ray derived gas model with a 20\% bias in cluster mass determination due to non-thermal pressure support, and BAO priors to derive $\sigma_8 = 0.77 \pm 0.02$. 
It has been pointed out by several authors \citep[e.g.,][]{hou14,wyman14,battye14} that discrepancies between low-redshift measurements of $\sigma_8$ and those based on primary CMB anisotropies can be reconciled by introducing the sum of the neutrino masses as a free parameter, with a value of $\summnu \sim 0.3 \ \mathrm{eV}$ preferred by the combination of data. 

The tension among measurements of tSZ power or tSZ-selected cluster counts, modeling of the tSZ effect, and the value of $\sigma_8$ preferred by primary CMB and BAO data appears to point to either inadequacies in our understanding of cluster physics, interesting new cosmological information, or some combination of the two.
Current cluster data and theoretical models are insufficient to disentangle the relative importance of these two effects. 
In the high-mass regime important for cluster count measurements, weak-lensing mass measurements of large, uniformly selected cluster samples (including out to high redshift with \emph{HST}) hold great promise for constraining the level of non-thermal pressure support and tSZ observable-mass scaling relations in general. 
In the lower-mass regime important for tSZ power spectrum measurements, we will have to rely in the near term on advances in theoretical modelling, combined with improved measurements of high-mass clusters and measurements such as the tSZ bispectrum that provide a bridge between the two regimes.

\section{kSZ interpretation} 
\label{sec:kszinterp}

\begin{figure*}[t]\centering
\includegraphics[width=0.9\textwidth,trim = 2.28cm 11.77cm 3.15cm 6.33cm]{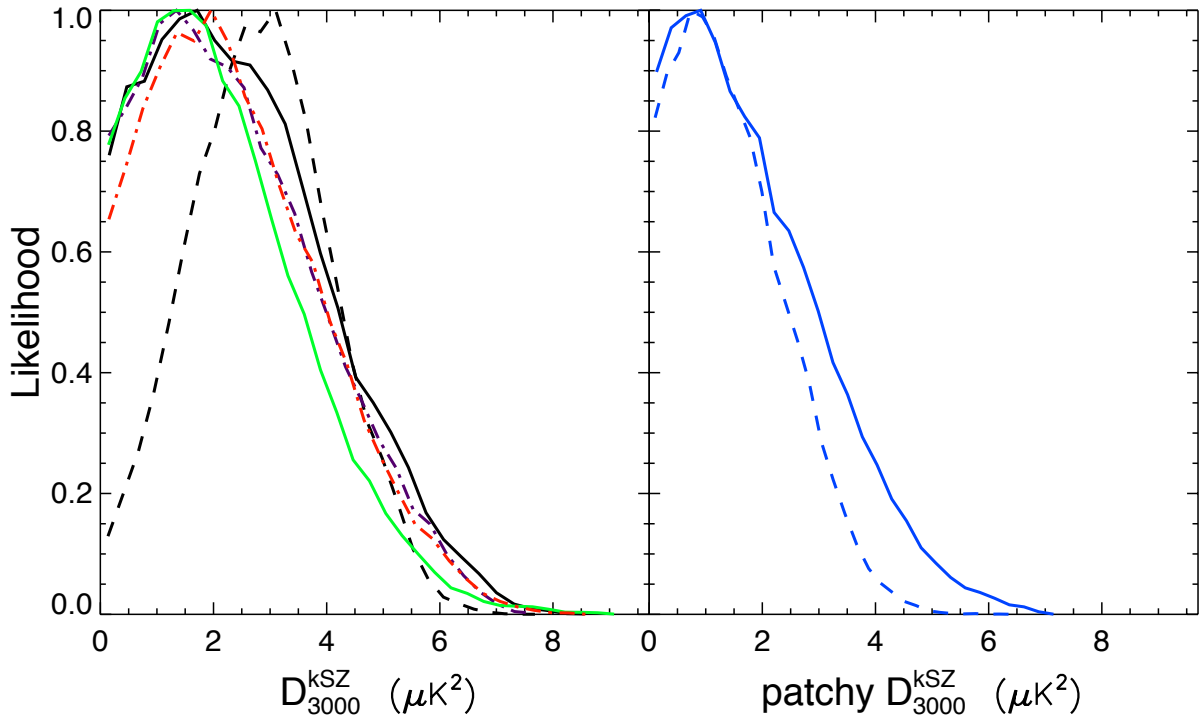}
  \caption[]{ 
    1D likelihood curves for the amplitude of the kSZ power spectrum at  $\ell=3000$. 
    \textbf{\textit{Left panel:}} The total kSZ amplitude, fit with a kSZ template with equal homogeneous and patchy kSZ contributions. 
    The \textbf{solid, black line} marks the fiducial model with the \textbf{dashed, black line} showing the impact of adding the bispectrum information. 
    The \textbf{dot-dashed, purple line} reflects the kSZ constraints when the 1- and 2-halo templates for the CIB clustering are used instead of the power-law approximation. 
    The \textbf{solid, green line} shows the constraint when a slope in the tSZ-CIB correlation as a function of angular multipole is introduced. 
    The \textbf{dot-dashed, red line} marks the model with the temperature allowed to vary in the modified BB model for the frequency scaling of the CIB. 
 \textbf{\textit{Right panel:}} Constraints on the patchy kSZ power with the CSF homogeneous kSZ model fixed.  Dashed and solid lines show results with and without included bispectrum information.
 }
  \label{fig:1dksz}
\end{figure*}

We now consider our measurements of the kSZ power spectrum and the implications for reionization. 
We  present constraints on the amplitude of the total kSZ power spectrum for a combined homogeneous + patchy kSZ template.
We also constrain the patchy kSZ power when fixing the homogeneous kSZ power to the CSF model.
Finally, we discuss the implications of the observed patchy kSZ power for the duration of the epoch of reionization.

\subsection{kSZ model comparison and constraints} 
\label{sec:kszmodel}

Using the power spectra presented in this work, we have significantly tightened the upper limits on the total kSZ power.  
Table \ref{tab:szconstraint} lists the constraints on the tSZ and kSZ powers with and without the bispectrum information using a variety of templates. 
With the default kSZ template, the 95\% CL upper limit on the kSZ power at $\ell = 3000$ is $\dksz{} < 5.4$\,\uksq, a factor of 1.25 lower than the limit reported by R12. 
The uncertainties were reduced by a factor of two: if the kSZ power likelihood had peaked at 0 (as in R12), the upper limit would have been 3.6\,\uksq. 
This limit is robust to the assumed tSZ templates (change is $<$\,4\%), but  depends somewhat on the choice of kSZ template. 

The total kSZ power will be the sum of the homogeneous and patchy kSZ power, which have slightly different shapes in $\ell$-space but an identical frequency dependance.
The default template assumes equal power at $\ell=3000$ from the homogeneous (CSF) and patchy kSZ templates. 
We also test the $\ell$-shape dependence of the inferred kSZ power by fitting with a kSZ template of either of the two extremes: 100\% patchy kSZ or 100\% homogeneous kSZ. 
The CSF homogeneous template leads to a more stringent upper limit of $\dksz{} < 4.5$\,\uksq, with the limit relaxing by 2.1\,\uksq{} to $\dksz{} < 6.6$\, \uksq\, with the patchy kSZ template. 
With the inclusion of the bispectrum information, however, the difference in upper limits between the two extremes is narrowed to 1.1\, \uksq, with the CSF homogeneous kSZ template resulting in an upper limit of $\dksz{} < 4.4$\,\uksq, and the patchy template giving an upper limit of $\dksz{} < 5.5$\, \uksq. 
The difference in total power constraints with the various kSZ templates suggests there is a small amount of information coming from the angular dependence of the kSZ signal.
Although the two components are highly degenerate in the current data, future surveys may have the power to disentangle the homogenous and patchy kSZ signals.  

We find only minor changes in the kSZ power constraints for the CIB models considered. 
The largest shift occurs when a slope is allowed in the tSZ-CIB correlation, which leads to a kSZ limit of  $\dksz{} < 3.7$\,\uksq. 
These limits are not in tension with our fiducial homogeneous kSZ template \citep[the CSF model of][]{shaw12} which predicts $\dksz = 1.6$\,\uksq, with the potential of additional kSZ power from the epoch of reionization.
  
We find a preference for positive total kSZ power with the inclusion of the tSZ bispectrum constraint. 
The bispectrum information both pushes the ML kSZ power higher and reduces the uncertainties. 
With the bispectrum constraint, we measure $D^{\rm kSZ}_{3000} = 2.9 \pm 1.3\, \uksq{}$ and disfavor zero kSZ power at 98.1\% CL. 
This is consistent at $<1\,\sigma$ with earlier measurements of the kSZ power, but with substantially reduced uncertainties. 
C14 found $D^{\rm kSZ}_{3000} = 2.6 \pm 1.8 \,\uksq{}$ using the same SPT bispectrum data and the R12 bandpowers, while \citet{addison12b} measured $5.3^{+2.2}_{-2.4} \,\uksq{}$ using a combination of \spitzer, \herschel, \planck, ACT, and SPT data. 
The data presented here provide the best measurement to date of the kSZ power.
  
\subsection{Implication for the epoch of reionization} 
\label{sec:kszpatchy}

We now interpret the measured kSZ power in light of the epoch of reionization. 
Using an estimate for the kSZ power that was generated during the epoch of reionization and marginalizing over a grid of reionization models which predict optical depth and kSZ power, we can obtain constraints on the universe's ionization history. 

\subsubsection{Patchy kSZ power}

To constrain reionization, it is necessary to determine how much kSZ power originated during reionization (i.e.,~patchy kSZ) versus in the late-time, fully ionized universe (homogeneous kSZ). 
We model the patchy kSZ power using a template from Z12 in which reionization started at $z=11$ and ended at $z=8$.  
While many models exist for the amplitude of the homogeneous kSZ power, \citet{shaw12} argue that the CSF model is a robust lower limit on the the hkSZ power; we therefore fix the hkSZ power to the prediction of the CSF model to obtain an upper limit on the patchy kSZ power. 
The contribution of the 30\% uncertainty on the amplitude of the CSF model to the total uncertainty of the kSZ patchy power is negligible.

The upper limits on the patchy kSZ power with and without the bispectrum information are shown in rows four and eight of Table~\ref{tab:szconstraint}. 
With no bispectrum information, we find a 95\% CL limit on the patchy kSZ power of $4.4\, \uksq$.
The upper limit falls by $1.1\,\uksq$ to $\dksz < 3.3\,\uksq$ when the bispectrum information is included. 

\subsubsection{Ionization history and the duration of reionization}

We follow the method of Z12 (summarized in Section \ref{sec:pksz}) to turn the observations of kSZ power (from SPT) and optical depth (from \wmap) into constraints on the ionization history. 
In brief, we build a grid of reionization models, each one of which predicts the ionization fraction as a function of redshift, kSZ power, and optical depth. 
This grid is importance sampled by the observational constraints on kSZ power and optical depth to place constraints on the ionization history and related metrics like the duration of reionization. 

\begin{figure*}[htb]\centering
\includegraphics[width=0.45\textwidth, clip=true, trim = 2.02cm  12.16cm  9.4cm  6.0cm]{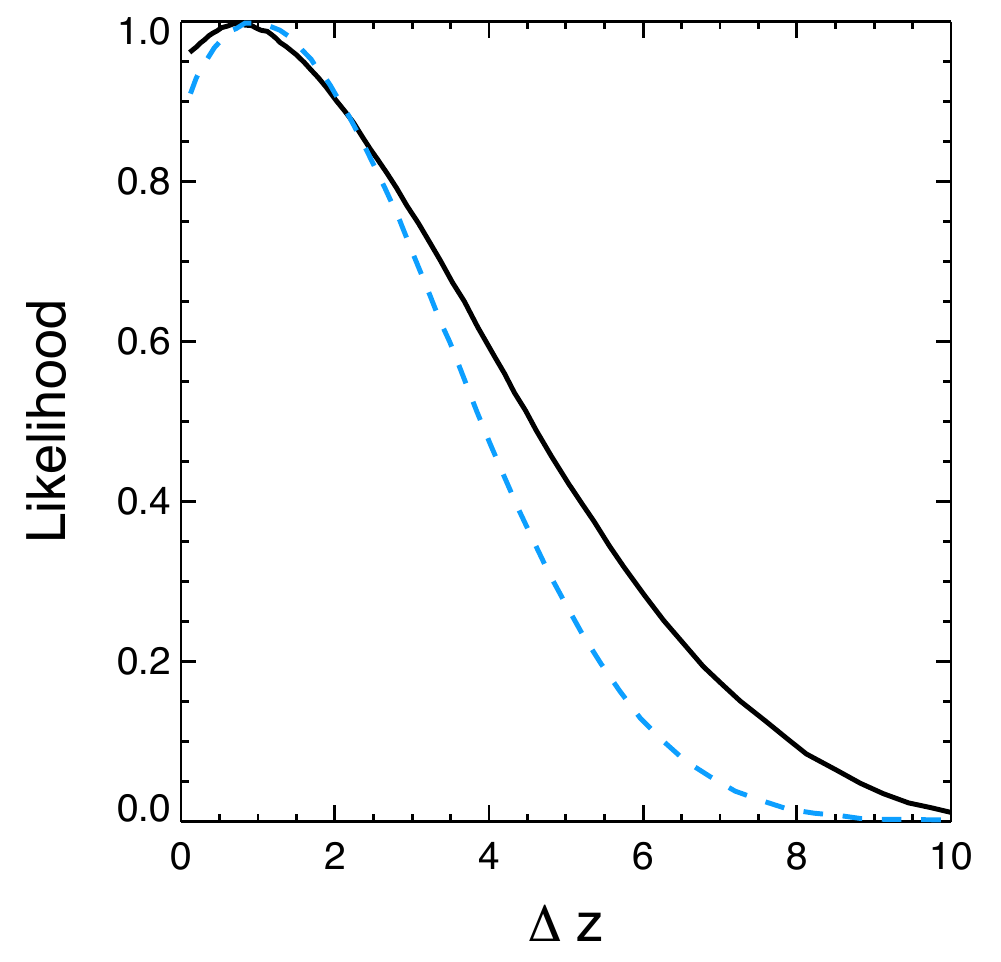}
\includegraphics[width=0.45\textwidth, clip=true, trim =2.02cm  12.16cm  9.4cm  6.0cm]{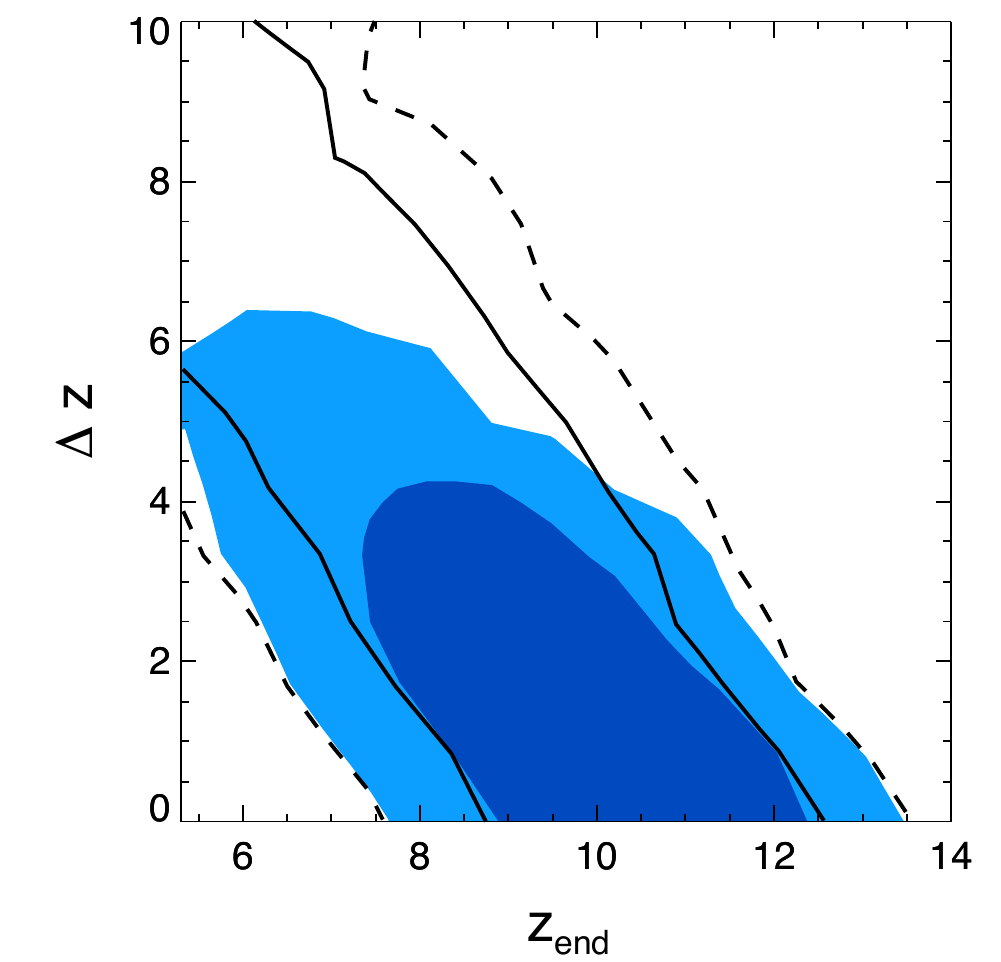}
  \caption[]{ 
  \textbf{\emph{Left:}} 1D likelihood curves for $\Delta z$ of reionization with the homogeneous kSZ fixed to the CSF model. 
  The \textbf{solid, black line} and  \textbf{dashed, blue line} lines show the constraint with and without including the bispectrum information.
  The model including bispectrum constraints leads to the tightest constraint on reionization with an upper limit on the duration $\Delta z < 5.4$ at 95\% CL with the likelihood peaking near $\Delta z = 1.3$.
   \textbf{\emph{Right:}} The \textbf{blue, filled contours} mark the 1 and 2 $\sigma$ contours of the 2D likelihood surface for $\Delta z$ and \zend\, ($x_e=0.99$), for the fixed CSF model including bispectrum information.
   Note that the width of the constraint on the \zend\, axis is primarily set by the errors on the WMAP measurement of $\tau$. The \wmap{} only constraint on reionization is shown with the \textbf{black, solid and dashed contours}.
 }
  \label{fig:dzplot}
\end{figure*}

The kSZ power amplitude primarily depends on the duration of reionization
(with the caveat that Z12 do not consider the effects of self-regulation, see Section \ref{sec:pksz}), 
while the timing is constrained by the optical depth. 
As this paper presents new kSZ constraints, we will focus on the duration of reionization. 
We define the duration as $\Delta z\equiv z_{\xhii=0.20}-z_{\xhii=0.99}$.
When including the bispectrum constraint, we find an upper limit on the duration $\Delta z < 5.4$  at 95\% CL.
The likelihood peaks around $\Delta z = 1.3$ in this case. 
 Figure \ref{fig:dzplot} shows the likelihoods for $\Delta z$ of reionization.

These measurements are consistent with the constraints presented in Z12 from the R12 dataset. 
In the comparable case in Z12 (a single parameter tSZ-CIB correlation with fixed CSF hkSZ), the authors found that $\Delta z < 7.9$ at 95\% confidence. 
Our data prefer a non-zero tSZ-CIB correlation, so we do not produce kSZ constraints with $\xi=0$, which produced the most stringent constraints on $\Delta z$ presented in Z12.
While the uncertainties with the current dataset are smaller than those presented in Z12, a more interesting difference is that with the current dataset the $\Delta z$ likelihood peaks above zero, driven by the combination of the more significant measurement of kSZ power presented here and the inclusion of the SPT bispectrum constraints from C14.


\section{tSZ-CIB correlation constraints} 
\label{sec:tszcibinterp}

We now examine constraints on the correlation between the galaxy clusters contributing to the tSZ power and the galaxies contributing to the CIB.
We consider two variations: (1) the baseline model where the correlation coefficient is independent of angular scale (see equation \ref{eqn:tszcib}), and (2) an alternative where the correlation depends linearly on angular multipole (equation \ref{eqn:tszcib-slope}). 
The latter model will be referred to as the ``sloped tSZ-CIB correlation" model. 
In all results quoted here, we tie the correlation coefficient at $\ell=3000$ to the tSZ-CIB correlation term in the \planck{} foreground model. 
DSFG over-densities in galaxy clusters lead to a positive correlation.

As in earlier works, we find an increase in kSZ power and a decrease in tSZ power with increasing correlation. 
As $\xi$ increases from 0.1 to 0.2, the tSZ power at 143 GHz drops by $\sim$1\,\uksq{} and the kSZ power rises by $\sim$2.5\,\uksq. 
Decreasing tSZ power with increasing correlation contradicts naive expectations. 
In single-frequency bandpowers (e.g.,~only 150\,GHz), increasing the  tSZ-CIB correlation would increase the allowed tSZ contribution for any observed power level. 
This naive picture changes when we consider the $95\times150$ and $150\times220$ bandpowers. 
The tSZ-CIB correlation reduces the power at $150\times150$ (which could be compensated by increasing tSZ power) but it has $\sim$50\% larger effect at $150\times220$ (where there is effectively zero tSZ power).
In contrast, a positive tSZ-CIB correlation reduces the power at $95\times150$ by a similar amount as in the $150\times150$ spectrum; however, there is more tSZ power in the 95\,GHz band.  
In the most sensitive combination of the three bands, adding kSZ power more effectively cancels the effects of positive tSZ-CIB correlation than adding tSZ power.

\begin{figure}[htb]
\begin{center}
\includegraphics[width=0.45\textwidth,clip=true,trim = 2.56cm 11.97cm 10.27cm 6.41cm]{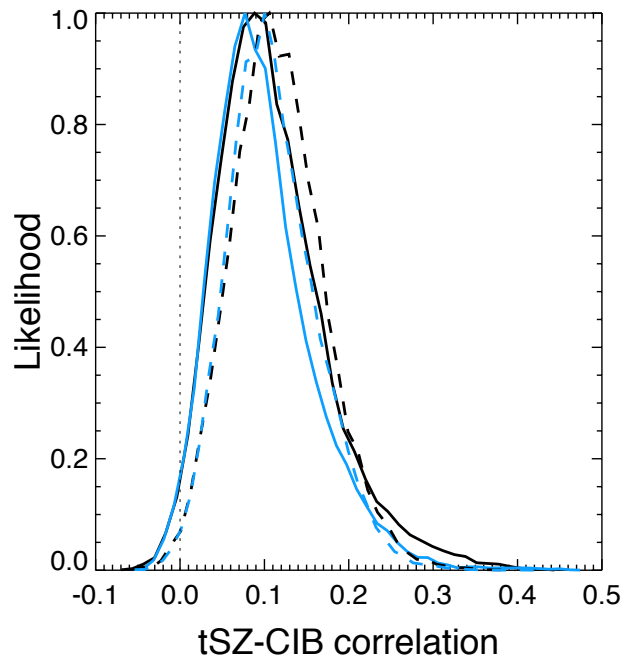}
\end{center}
\caption[ 
1d likelihood of tSZ-CIB correlation with and without bispectrum for 2 models]{
The \textbf{black solid line} shows the constraints on the tSZ-CIB correlation in our fiducial model. 
The \textbf{blue solid line} shows constraints on the tSZ-CIB correlation in a model where the angular slope of the correlation is a free parameter. 
The \textbf{dashed lines} are these same models including the prior on the tSZ power from the bispectrum measurement in C14. 
}
\label{fig:1paneltszcib}
\end{figure}

In the baseline model, we find a preferred tSZ-CIB correlation of $\xi = 0.100^{+0.069}_{-0.053}$, 
with negative correlation ($\xi <0$) ruled out at 98.1\% CL. 
This value is consistent with the R12 reported value of $\xi = 0.18 \pm 0.12$, but with substantially reduced uncertainty.  
In contrast to R12, the distribution we see here has no long tail toward large positive values of correlation. 
Such a tail arose in the R12 analysis when tSZ power approached zero, at which point the exact degree of  correlation ceased to matter.
The more significant detection of tSZ power in this work excludes this edge condition. 

The preferred correlation at $\ell=3000$ is essentially unchanged in the sloped tSZ-CIB correlation model: $\xi_{3000} = 0.090^{+0.064}_{-0.048}$. 
We have no evidence for a slope in the tSZ-CIB correlation.
The SZ results are insensitive to a slope in the correlation coefficient in the prior range of [-1, 1].

Including a prior on the tSZ power from the bispectrum measurement increases the preferred value of the tSZ-CIB correlation roughly 0.5\,$\sigma$ in both models considered. 
Including bispectrum constraints, in the fiducial model $\xi = 0.113^{+0.057}_{-0.051}$, and in the model with the angular slope of $\xi$ left free, $\xi = 0.105^{+0.057}_{-0.046}$. 
Figure \ref{fig:1paneltszcib} shows the constraints for both models considered with and without the bispectrum constraint.

\section{Cosmic infrared background constraints} 
\label{sec:cibinterp}

Finally, we examine the power constraints on the cosmic infrared background. 
 SPT data strongly constrain the CIB power spectrum at millimeter wavelengths, especially at 220\,GHz. 
 The CIB constraints for a variety of model assumptions are summarized in Table \ref{tab:cibparams}. 
Recall that sources detected above $5 \sigma$ at 150\, GHz ($\sim$6.4\,mJy) have been masked. 
For this flux cut, the Poisson CIB term is slightly dominant at $\ell = 3000$ and 220\,GHz in all models considered, accounting for $\sim$55\% of the total CIB power. 
The Poisson term increases in importance towards lower frequencies contributing 70\% of the total power at 150\,GHz and 85\% at 95\,GHz.

In the baseline model, we find the 220\,GHz Poisson power to be $D^{p}_{3000}=66.1 \pm 3.1\, \mu {\rm K}^2$. 
The clustered amplitude is $D^{c}_{3000}=50.6 \pm 4.4\, \mu {\rm K}^2$.  
At 150\,GHz, the Poisson and clustering powers are $D^{p}_{3000}=9.16 \pm 0.36\, \mu {\rm K}^2$ and $D^{c}_{3000}=3.46 \pm 0.54\, \mu {\rm K}^2$ respectively. 
The estimated CIB uncertainties at 95\,GHz are very small, but we caution that this extrapolation from higher frequencies is very sensitive to the frequency modeling. 
Both Poisson and clustered CIB components have more power at 150\,GHz than the SZ spectra across the multipoles ($\ell > 2000$) important to the SZ measurement.

Compared to R12, the uncertainties in the powers are a factor of 10-20\% smaller despite having more degrees of freedom in the foreground model. 
As noted in Section \ref{sec:cibbaseres}, the primary results presented in R12 use a different foreground model. 
When the R12 bandpowers are used with the model presented in this work, we find $D^{p}_{3000}= 8.59 \pm0.63$\,\uksq{} and $D^{c}_{3000} = 4.00\pm1.1$\,\uksq{}  at 150 GHz, which represents a shift of less than 1$\sigma$ between  the two data sets.
We measure the same total CIB power with the two models, but with the increased freedom in the foreground model a higher fraction of the power at 150GHz is attributed to the Poisson term rather than the clustered term.

As in R12, the data prefer a steep drop-off in CIB power from 220 to 150\,GHz; the Poisson spectral index is $\alpha^{p}_{150/220} = 3.267 \pm 0.077$ and the clustered spectral index is $\alpha^{c}_{150/220} = 4.27 \pm 0.20$.  
The spectral indices do not depend on shape of the clustered term, tSZ-CIB correlation, or allowing $T$ to vary (see Table \ref{tab:cibparams}). 
R12 showed a spread in spectral index in the same direction, which is consistent with this data when using the new foreground model. 
Some difference between the Poisson and clustered spectral indices is expected due to the different redshift dependences of the contributions to each of the terms. 

The observed clustered spectral index is $2.8\,\sigma$ higher than the $\alpha_{150/220}^{c} = 3.68 \pm 0.07$ ($\beta = 2.2 \pm 0.07$) that was found in an analysis of ACT and BLAST data \citep{addison12a} (but consistent with R12, once the new foreground model is taken into account). 
The \citet{addison12a} analysis, however, spans higher frequencies (up to $\sim850$\,GHz).
As discussed in \citep{meny07}, the emission of both amorphous and crystalline dust grains is both temperature and frequency dependent.  
\citet{boudet05} show a trend towards higher $\beta$ at lower frequencies in laboratory measurements of sample dust grains.
Additionally, \citet{paradis09} showed that $\beta$ will vary with frequency when the dust is multi-component, but described with a single isothermal mode.
Differing spectral coverage might account for the  differences in $\alpha_{150/220}^{c}$.

The two model extensions that alter the shape of the clustered term show modest improvements to the quality of fit to the data (\delchisq = -1.8 for 1 dof in both cases). 
The SZ constraints are not affected in either case. 
The first extension introduces separate 1 and 2-halo clustered templates with free amplitudes (but the same frequency scaling). 
The relative power in the Poisson and clustered terms is essentially unchanged, with a less than 1$\sigma$ shift in power. 
The second extension allows the power-law exponent of the clustered term to vary; there is a mild preference for an index below 0.8. 

\begin{figure*}[tbh]\centering
\includegraphics[trim=2cm 12.7cm 1cm 7cm,clip,width=0.9\textwidth]{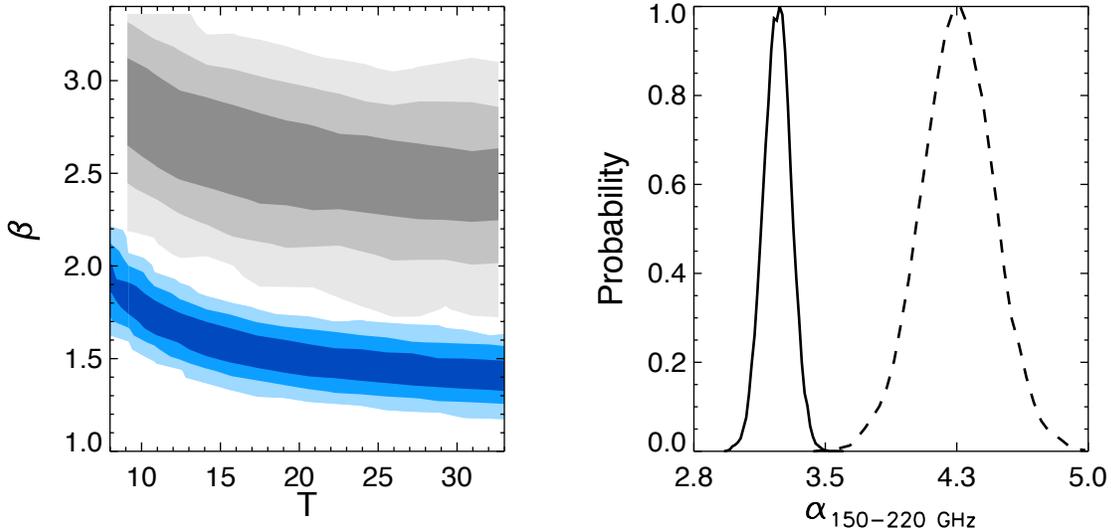}
  \caption[]{ 
  CIB frequency modeling. 
  \textit{\textbf{Left panel:}} 
  2D likelihood surface for T and $\beta$ for the modified BB CIB spectrum for the Poisson (\textbf{blue contours}) and clustered (\textbf{grey contours}) terms. 
  \textit{\textbf{Right panel:}}
  The effective CIB spectral index between 150 and 220\,GHz for the Poisson (\textbf{solid line}) and clustered (\textbf{dashed line}) terms.
  }
  \vspace{1.0cm}
  \label{fig:cibfreqconstraints}
\end{figure*}

As shown in Figure~\ref{fig:cibfreqconstraints}, there is a clear degeneracy between the two parameters of the modified BB model with only SPT data. 
Recall that in this model, the CIB spectrum is described by $\nu^\beta B_\nu(T)$. 
Note that temperatures above $\sim$\,$20\,$K are essentially indistiguishable since all three observing frequencies are in the Rayleigh-Jeans region of the spectrum. 
There is a preference for higher temperatures (not shown) when the temperature is required to be the same for both the Poisson and clustered CIB terms.

\begin{table*}[ht!]
\begin{center}
\caption{\label{tab:cibparams} CIB Constraints}
\small
\begin{tabular}{l|ccc|ccc}
\hline\hline
\rule[-2mm]{0mm}{6mm}
Model &  \multicolumn{3}{c}{Poisson ($\mu {\rm K}^2$)} & $\alpha_p^{150-220}$ & $\alpha_c^{150-220}$ & \\
& 95\,GHz & 150\,GHz & 220\,GHz & &&\\
\hline
Baseline   & $ 1.37 \pm  0.13$ & $ 9.16 \pm 0.36$ & $ 66.1 \pm  3.1$ & $3.267 \pm 0.077$ & $ 4.27 \pm  0.20$ \\  
~~1/2 halo cluster model   & $ 1.27 \pm  0.14$ & $ 8.61 \pm 0.78$ & $ 62.6 \pm  9.4$ & $3.282 \pm 0.126$ & $ 4.19 \pm  0.21$ \\  
~~tSZ-CIB slope   & $ 1.40 \pm  0.20$ & $ 9.22 \pm 0.54$ & $ 65.9 \pm  3.0$ & $3.250 \pm 0.099$ & $ 4.30 \pm  0.21$ \\  
~~T varying   & $ 1.35 \pm  0.15$ & $ 9.14 \pm 0.35$ & $ 66.0 \pm  3.0$ & $3.264 \pm 0.076$ & $ 4.31 \pm  0.20$ \\  
\hline\hline
 & \multicolumn{3}{c}{$\ell^{0.8}$ Clustering ($\mu {\rm K}^2$)}& \multicolumn{3}{c}{Lin. Theory Clustering ($\mu {\rm K}^2$)} \\
& 95\,GHz & 150\,GHz & 220\,GHz & 95\,GHz & 150\,GHz & 220\,GHz  \\
\hline
Baseline   & $0.208 \pm 0.068$ & $ 3.46 \pm 0.54$ & $ 50.6 \pm  4.4$ & - & - & - \\ 
~~1/2 halo cluster model   & $0.122 \pm 0.087$ & $ 1.90 \pm 1.06$ & $ 26.5 \pm 12.5$ & $0.128 \pm 0.040$ & $ 1.99 \pm 0.35$ & $ 28.3 \pm  4.5$ \\ 
~~tSZ-CIB slope   & $0.201 \pm 0.068$ & $ 3.43 \pm 0.55$ & $ 51.1 \pm  4.3$ & - & - & - \\ 
~~T varying   & $0.192 \pm 0.067$ & $ 3.39 \pm 0.53$ & $ 50.8 \pm  4.2$ & - & - & - \\ 

\hline
\end{tabular}
\tablecomments{  
Constraints on the CIB power and spectral index are presented for four CIB models. 
All of these models use the modified BB frequency dependence for the CIB. 
The first row is for the baseline model. 
The next three rows consider parameter extensions to the baseline CIB treatment. 
In the ``tSZ-CIB slope" row, we allow the tSZ-CIB correlation to have a linear slope in $\ell$. 
In the ``1-2 halo'' row, we use the 1-2 halo model instead of the $\ell^{0.8}$ spatial template for the clustered CIB.   
In the ``T varying" row, the temperature of the modified BB CIB frequency scaling is allowed to vary between 5 and 50\,K.  
For each model, we present the constraints on the amplitudes in $\mu{\rm K}^2$ at $\ell = 3000$ of the Poisson and  clustering CIB templates. 
The right column shoes the effective spectral indices from 150 to 220\,GHz. 
  }
\normalsize
\end{center}
\end{table*}

\section{Conclusions} 
\label{sec:conclusions}
We have presented high angular resolution temperature power spectra from the complete $2540$\,\sqdeg{} \sptsz{} survey, observed between  2008 and 2011.
The survey area, comprising 6\% of the total sky, has been mapped to depths of approximately 40, 18, and $70\,\mu$K-arcmin at 95, 150, and $220\,$GHz respectively. 
The six auto- and cross-spectra from observations at these three frequencies are shown in
Table~\ref{tab:bandpowers} for $\ell \in [2000,11000]$.  
The bandpowers  measure arcminute-scale millimeter wavelength anisotropy. 
In addition to the primary CMB temperature anisotropy, significant power is contributed by DSFGs, radio galaxies, and the kinematic and thermal SZ effects.

We perform multi-frequency fits using MCMC methods to a combined data set including \planck{}, \wmap{} polarization, BAO, H$_0$, and the  SPT bandpowers presented in this work. 
We find that the minimal model  to explain the current data adds seven free parameters beyond the six $\Lambda$CDM parameters: two describing the Poisson component of the DSFGs, two describing the clustered component of the DSFGs, one describing the amplitude of the tSZ power spectrum, one describing the correlation between the tSZ and DSFGs, and one describing the spectral index of radio galaxies. 
Strong priors are included on the amplitude of the radio galaxies and galactic cirrus. 
Although not required to fit the data, we also include a free parameter for the amplitude of the kSZ power spectrum. 
The SPT bandpowers are fit well by this eight parameter model with a PTE of 21.5\%. 
We explore a number of extensions to this baseline model as well as alternate template shapes.
These variants do not significantly improve the quality of fits to the data, although in some cases, they slightly alter the conclusions drawn. 

We fit for the amplitude of the tSZ power spectrum in the baseline model and find  $D^{\rm tSZ}_{3000}=4.38^{+0.83}_{-1.04}\, \mu {\rm K}^2$  at 143\,GHz.
For the three models of the tSZ power considered (Shaw, Sehgal, and Bhattacharya), the constraint on the tSZ power is independent of the tSZ template used when fitting.
When including a constraint on the tSZ power from the 800 deg$^2$ SPT bispectrum, we find that the constraint on the tSZ power tightens to $D^{\rm tSZ}_{3000} = 4.08^{+0.58}_{-0.67}\,\uksq{}$. 

We consider whether the measurement of tSZ power can improve constraints on $\sigma_8$ in the standard cosmological model, and find little ($<$\,10\%) improvement assuming 50\% modeling uncertainty.
We find some tension, however, between the measurements of tSZ power presented here and the level predicted by models informed by the value of $\sigma_8$ preferred by primary CMB and BAO data. 
This tension could be due to inadequacies in our current understanding of cluster physics or to a non-standard value for some cosmological parameter such as the sum of the neutrino masses. 
Improvements in cluster data expected from upcoming weak-lensing measurements and more precise measurements of the tSZ bispectrum, combined with advances in cluster modeling, could help break this degeneracy.

We find a 2.2$\sigma$ preference for positive total kSZ power when the tSZ bispectrum is included, finding $D^{\rm kSZ}_{3000} = 2.9 \pm 1.3 \uksq{}$,
 when fitting with a template with equal power at $\ell=3000$ from the CSF homogenous and patchy kSZ models.  
Negative kSZ power is disfavored at 98.1\% CL. 
The 95\% CL upper limit is $D^{\rm kSZ}_{3000} < 4.9\,\uksq{}$. 
When fixing the homogeneous kSZ to the CSF model and including the bispectrum information, we constrain the amplitude of the patchy kSZ to be $D^{\rm pkSZ}_{3000} < 3.3\,\uksq{}$.

We use the SPT kSZ power and the \wmap{} optical depth measurements to constrain the duration and end of reionization. 
Assuming the CSF model predictions for the homogeneous kSZ power, we set an upper limit on the duration $\Delta z < 5.4$ at 95\% CL with the likelihood peaking near $\Delta z \sim 1.3$.

We find that the data prefer a positive tSZ-CIB correlation with $\xi = 0.100_{-0.053}^{+0.069}$. 
Positive correlation is favored at a confidence level of 98.1\%. 
The preferred value of the correlation shifts to $\xi = 0.113_{-0.051}^{+0.057}$ with the inclusion of the bispectrum constraint on the tSZ power.
We also investigate an $\ell$-dependence in the tSZ-CIB correlation, finding an extremely weak preference for a negative slope and no significant impact on the SZ constraints. 
As the tSZ-CIB correlation is degenerate with the tSZ and kSZ powers, external data or models to constrain this correlation would significantly tighten constraints on the tSZ and kSZ powers.

We test whether allowing more freedom in the primary CMB temperature power spectrum, such as including significant tensors ($r=0.2$), changing the number of neutrino species, or allowing the neutrino mass to be a free parameter, affects the SZ constraints and find that it does not. 
We also investigate whether alternative CIB modeling assumptions lead to changes in the SZ constraints, and do not find a dependence. 

The SPT data provide constraints on the cosmic infrared background. 
We fit separately for the amplitude and spectral index of the Poisson and clustered components of the DSFGs, resulting in a 4 parameter model to describe the CIB emission.
When adjusting for the model change between R12 and this work, the measured CIB power agrees well with previous estimates by R12. 
The fluctuation power of the DSFGs falls sharply from 220 to $150\,$GHz, with effective spectral indices for the Poisson and clustered components of  $\alpha^{\rm 150-220}_p = 3.267 \pm 0.077$  and $\alpha^{\rm 150-220}_c = 4.27 \pm 0.20$.
The preferred clustered spectral index is higher than reported by \citet{addison12a}, which might be explained by the frequency ranges used in each work.

The SPT is currently observing 500\,\sqdeg{} with the SPTpol camera which will result in substantially lower noise at 95 and 150\,GHz \citep{austermann12}.
In addition, 100\,\sqdeg{} of the survey was observed by \herschel/SPIRE in 2012, and analysis of the combined SPT and \herschel\ dataset is ongoing. 
The combined data, with 21 frequency cross spectra, will be a valuable resource in understanding the CIB and its correlation with the thermal SZ signal, and will result in significantly tighter constraints on the reionization history of the Universe.

\acknowledgments

The South Pole Telescope is supported by the National Science Foundation through grant PLR-1248097.  Partial support is also provided by the NSF Physics Frontier Center grant PHY-1125897 to the Kavli Institute of Cosmological Physics at the University of Chicago, the Kavli Foundation and the Gordon and Betty Moore Foundation grant GBMF 947.
The McGill group acknowledges funding from the National Sciences and Engineering Research Council of Canada, Canada Research Chairs program, and the Canadian Institute for Advanced Research. 
R. Keisler acknowledges support from NASA Hubble Fellowship grant HF-51275.01.
M. Dobbs acknowledges support from an Alfred P. Sloan Research Fellowship.
This research used resources of the National Energy Research Scientific Computing Center, which is supported by the Office of Science of the U.S. Department of Energy under Contract No. DE-AC02-05CH11231. 
We acknowledge the use of the Legacy Archive for Microwave Background Data Analysis (LAMBDA). Support for LAMBDA is provided by the NASA Office of Space Science.

\bibliography{highell2500}

\end{document}